\documentclass[seceq,preprint]{ptptex}

\usepackage{graphicx}
\usepackage{verbatim}

\preprintnumber[2cm]{
YGHP-12-50}

\newcommand{\refer}[1]{(\ref{#1})}
\newcommand{\oper}[1]{\mathcal{#1}}
\newcommand{\Tr}[1]{\mathrm{Tr}\left(#1\right)}

\newcommand{\comm}[2]{\left[ #1, #2 \right]}
\newcommand{\mean}[1]{\left\langle #1\right\rangle}

\def\beq{\begin{eqnarray}}
\def\eeq{\end{eqnarray}}
\def\Tr{{\rm Tr}}
\def\Lag{\mathcal{L}}
\def\p{\partial}
\def\D{\mathcal{D}}

\newcommand {\non}{\nonumber\\}

\title{
Matter Fields and Non-Abelian Gauge Fields Localized on Walls
}

\author{
Masato \textsc{Arai}$^{a,b,}$\footnote{E-mail: masato.arai@gmail.com} 
Filip \textsc{Blaschke}$^{b,c,}$\footnote{E-mail: filip.blaschke@fpf.slu.cz}
Minoru \textsc{Eto}$^{d,}$\footnote{E-mail: meto@sci.kj.yamagata-u.ac.jp} 
and Norisuke \textsc{Sakai}$^{e,}$\footnote{E-mail: norisuke.sakai@gmail.com
}
}

\inst{
$^a$Fukushima National College of Technology, Iwaki, Fukushima 970-8034, Japan\\
$^b$Institute of Experimental and Applied Physics, Czech Technical University in Prague, Horsk\'a 22, 128 00 Prague 2, Czech Republic\\
$^c$Institute of Physics, Silesian University in Opava, Bezru\v{c}ovo n\'am. 1150/13, 746~01 Opava, Czech Republic\\
$^d$Department of Physics, Yamagata University, Yamagata 990-8560, Japan\\
$^e$Department of Mathematics, Tokyo Woman's Christian University, \\
Tokyo 167-8585, Japan
}

\abst{
}

\abst{%
Massless matter fields and non-Abelian gauge fields are 
localized on domain walls in a (4+1)-dimensional $U(N)_c$ 
gauge theory with $SU(N)_{L}\times SU(N)_{R}\times U(1)_{A}$ 
flavor symmetry. 
We also introduce $SU(N)_{L+R}$ flavor gauge fields 
and a scalar-field-dependent gauge coupling, 
which provides massless non-Abelian gauge fields localized 
on the wall. 
We find a chiral Lagrangian interacting minimally with the 
non-Abelian gauge field together with nonlinear 
interactions of moduli fields as the (3+1)-dimensional 
effective field theory  
up to the second order of derivatives. 
Our result provides a step towards a realistic model building 
of brane-world scenario using topological solitons. 
}

\begin{document}

\maketitle

\section{Introduction}\label{sc:introduction}

Gauge hierarchy problem is a good guiding principle to 
construct theories beyond the Standard Model (SM). 
Brane world scenario \cite{Horava:1995qa,ArDiDv,RaSu} 
is one of the most attractive proposals to solve this 
problem, besides models with supersymmetry (SUSY) 
\cite{Dimopoulos:1981zb}. 
In the brane world scenario, it is assumed that all fields 
except the graviton field are localized on (3+1)-dimensional 
world volume of a defect called 3-brane, immersed in a 
many-dimensional space\discretionary{-}{-}{-}time 
called bulk.
In order to realize such a scenario dynamically, 
we may use a topological soliton. 
For instance, let us consider a domain wall solution as 
the simplest soliton. 
To obtain (3+1)-dimensional world volume on the domain 
wall, we need to consider a theory in a (4+1)-dimensional 
space-time. 
Bulk fields in (4+1)-dimensions can provide massless 
modes localized on the domain wall, besides many massive 
modes in general. 
After integrating over massive modes, one obtains 
 low-energy effective field theory describing 
the effective interactions of massless modes. 
Massless matter fields have been successfully localized 
on domain walls \cite{RuSh}, 
but localization of the gauge field on domain walls in field theories has been difficult \cite{DuRu}. 
It has been noted that the broken gauge symmetry 
in the bulk outside of the soliton inevitably makes 
the localized gauge field massive with the mass of the 
order of inverse width of the wall\cite{DaSh,Maru:2003mx}. 
To localize a massless gauge field, one needs to have 
the confining phase rather than the Higgs phase in the bulk 
outside of the soliton. 
Earlier attempts used a tensor multiplet in order to implement 
Higgs phase in the dual picture, but this approach successfully  
localize only $U(1)$ gauge field \cite{IsOhSa}. 
More recently, a classical realization of the 
confinement \cite{Kogut:1974sn,Fukuda:1977wj} through the 
position-dependent gauge coupling 
has been successfully applied to localize the non-Abelian 
gauge field on domain walls \cite{OhSa}. 
The nontrivial profile of this position-dependent 
gauge coupling was naturally introduced on the domain wall 
background through a scalar-field-dependent gauge coupling 
function resulting from a cubic prepotential of supersymmetric 
gauge theories. 
The appropriate profile of the position-dependent 
gauge coupling was obtained from domain wall solutions 
using two copies of the simplest model or 
from a model with less fields and a particular mass 
assignment. 
However, it was still a challenge to introduce matter 
fields in nontrivial representations of the gauge group 
of the localized gauge field. 

Parameters of soliton solutions are called moduli 
and can be promoted to fields on the world volume 
of the soliton. 
Massless fields in the 
low-energy effective field theory on the soliton 
background are generally given by these moduli fields. 
Moduli with non-Abelian global symmetry is often called 
the non-Abelian cloud, and has been explicitly realized 
in the case of domain walls using Higgs scalar fields 
with degenerate masses in $U(N)_c$ gauge theories 
\cite{EtFuNiOhSa}. 
This model also has a non-Abelian global symmetry 
$SU(N)_L\times SU(N)_R\times U(1)_{A}$, which is 
somewhat similar to the chiral symmetry of QCD. 
If we turn this global symmetry into a local gauge 
symmetry, we should be able to obtain the usual minimal gauge 
coupling between these moduli fields and the gauge field. 
Since we wish to localize the gauge field 
on the domain wall, it is essential to choose the 
global symmetry of moduli fields to be unbroken in 
the vacua (of both left and right bulk outside of the wall). 
This choice will guarantee that the bulk 
outside of the domain wall is not in the Higgs phase. 
Therefore we are led to an idea where we introduce 
gauge fields corresponding to a flavor symmetry group 
of scalar fields which will be unbroken in the vacuum. 
If we introduce the additional scalar-field-dependent 
gauge coupling function similarly to the supersymmetric model, 
we should be able to localize both 
massless matter fields and the massless gauge field 
at the same time on the domain wall.

The purpose of this paper is to present a (4+1)-dimensional 
field theory model of localized massless matter fields 
minimally coupled to the non-Abelian gauge field which 
is also localized on the domain wall with the 
(3+1)-dimensional world volume. 
We also derive the low-energy effective field theory of 
these localized matter and gauge fields. 
To introduce non-Abelian flavor symmetry (to be gauged 
eventually) in the domain wall sector, we replace 
one of the two copies of the $U(1)_c$ gauge theory 
with the flavor symmetry $U(1)_L\times U(1)_R$ in 
Ref.\citen{EtFuNiOhSa}, by $U(N)_c$ gauge theory with 
the extended flavor (global) symmetry 
$SU(N)_L\times SU(N)_R\times U(1)_{A}$. 
By choosing the coincident domain wall solution for this 
domain wall sector, we obtain the 
maximal unbroken non-Abelian flavor symmetry group 
$SU(N)_{L+R}$ which is preserved in both left and 
right vacua outside of the domain wall. 
Therefore we can introduce gauge field for the 
(subgroup of) the flavor $SU(N)_{L+R}$ symmetry. 
In order to obtain the field-dependent gauge coupling 
function, for the gauge field localization mechanism 
 \cite{OhSa}, we also introduce a coupling between 
a scalar field and gauge field strengths 
inspired by supersymmetric gauge theories, 
although we do not make the model fully supersymmetric 
at present. 
This scalar-field-dependent gauge coupling function gives 
appropriate profile of position-dependent 
gauge coupling through the background domain wall solution. 
With this localization mechanism for gauge field, we find 
massless non-Abelian gauge fields localized on the domain 
wall. 
We also obtain the low-energy effective field theory 
describing the massless matter fields in the non-trivial 
representation of non-Abelian gauge symmetry. 
Since our flavor symmetry resembles the chiral symmetry 
of QCD before introducing the gauge fields that are 
localized, we naturally obtain a kind of chiral Lagrangian 
as the effective field theory on the domain wall. 
We find an explicit form of full nonlinear interactions of 
moduli fields up to the second order of derivatives. 
Moreover, these moduli fields are found to interact with 
$SU(N)_{L+R}$ flavor gauge fields as adjoint 
representations. 
In analyzing the model, we use mostly the strong coupling 
limit for the domain wall sector. 
The strong coupling is merely to describe 
our result explicitly at every stage. 
Even if we do not use the strong coupling, 
the physical features are unchanged. 
It is easy to expect that (the part of) the gauge 
symmetry is broken when the walls separate in 
each copy of the domain wall sector. 
Our results of the low-energy effective field theories 
shows that flavor gauge symmetry $SU(N)_{L+R}$ is 
broken on the non-coincident wall 
and the associated gauge bosons acquire masses 
as walls separate. 
This geometrical Higgs mechanism is quite similar to D-brane systems
in superstring theory. So our domain wall system provides a genuine prototype
of field theoretical D3-branes. This is
an interesting problem, which we plan to analyze more in future. 
We also find indications that additional moduli will appear 
in the supersymmetric version of our model, which is also 
an interesting future problem to study. 

The organization of the paper is as follows. 
In section \ref{sec:Abelian-Higgs}, 
we explain the localization mechanism by taking 
Abelian gauge theory as an illustrative example. 
In section \ref{sec:chiral model}, we introduce the chiral model with the 
non-Abelian flavor symmetry for the domain wall sector 
and then also introduce gauge fields for the unbroken part of 
the flavor symmetry. 
By introducing the scalar-field-dependent gauge coupling 
function, 
we arrive at the localized massless gauge field interacting 
with the massless matter field in a nontrivial 
representation of flavor gauge group. 
The low-energy effective field theory is also 
worked out. 
In section \ref{sec:SUSYE}, an attempt is made to make the model 
supersymmetric. 
New additional features of the supersymmetric models 
are also described. 
In section \ref{sc:6}, 
we summarize our results and discuss remaining 
issues and future directions. 
In Appendix \ref{app3} we discuss domain wall solution for 
gauged massive $\mathbb{C}P^1$ sigma model.
Appendix \ref{app2} describes derivation of effective 
Lagrangian which includes full nonlinear 
interactions between moduli fields.
Appendix \ref{app:det} contains derivation of positivity 
condition for the potential appearing in section \ref{sec:SUSYE}.

\bigskip
\section{Abelian-Higgs model 
of gauge field localization
}
\label{sec:Abelian-Higgs}

\subsection{The domain wall sector}

Let us illustrate the localization mechanism for the gauge 
fields and the matter fields on the domain walls 
by using a simplest model in (4+1)-dimensional 
spacetime : two copies ($i=1, 2$) 
of $U(1)$ models, each of which has two flavors 
($L, R$) of charged Higgs scalar fields 
$H_i=(H_{iL}, H_{iR})$ :
\beq
\Lag_i &=& - \frac{1}{4g_i^2}\left({\cal F}_{MN}^i\right)^2 
+ \frac{1}{2g_i^2}\left(\p_M\sigma_i\right)^2 
+ \left|\D_M H_i\right|^2 - V_i, \label{Abelian-Higgs} \\
V_i &=& \frac{g_i^2}{2}\left(|H_i|^2 - v_i^2\right)^2 
+ \left|\sigma_iH_i-H_iM_i\right|^2. 
\eeq
We use the metric $\eta_{MN}={\rm diag}(+,-,\cdots,-)$, 
$M,N=0,1,\cdots,4$. 
The Higgs field $H_i$ is charged with respect to the $U(1)_i$ 
gauge symmetry and the covariant derivative is given by
\beq
\D_M H_i = \p_M H_i + i w_M^i H_i,
\eeq
where $w_M^i$ is the $U(1)_i$ gauge field with the field strength 
\beq
{\cal F}_{MN}^i = \p_M w_N^i - \p_N w_M^i.
\eeq
Since we want domain walls, we will choose 
\beq
M_i = {\rm diag}\left(m_i,-m_i\right),\quad (m_i > 0),
\eeq
resulting in the $U(1)_{iA}$ flavor symmetry\footnote{
Phase rotation of $H_{iL}$ and $H_{iR}$ in the same 
direction $U(1)_i$ is gauged and the remaining global 
symmetry is in the opposite direction and is denoted as 
$U(1)_{iA}$. 
}. 
We have included the neutral scalar fields $\sigma_i$ in this 
Abelian-Higgs model. 
The gauge coupling $g_i$ appears not only in front of the 
kinetic terms of the gauge fields and $\sigma_i$, 
but also as the the quartic coupling constant of $H_i$. 
Both these features are motivated by the supersymmetry. 
Indeed, we can embed this bosonic Lagrangian into a 
supersymmetric model with eight supercharges by adding 
appropriate fermions and bosons, which will not play a role 
to obtain domain wall solutions. 
We have taken this special relation among the coupling 
constants only to simplify concrete computations below. 
One may repeat the following procedure in models with more 
generic coupling constants without changing essential results.

The first term of the potential is the wine-bottle type 
and the Higgs fields develop non-zero vacuum expectation 
values. There are two discrete vacua for each copy $i$ 
\beq
(H_{iL},H_{iR},\sigma_i) = \left(v_i,0,m_i\right),\ 
\left(0,v_i,-m_i\right).
\label{eq:vacua_AH}
\eeq

Thanks to the special choice of the coupling constants 
in $\Lag_i$ motivated by the supersymmetry, 
there are Bogomol'nyi-Prasad-Sommerfield (BPS) domain wall solutions in these models. 
Let $y$ be the coordinate of the direction orthogonal 
to the domain wall and we assume all the field depend 
on only $y$. 
Then, as usual, the Hamiltonian can be written as follows 
\beq
{\cal H}_i &=& \frac{1}{2g_i^2}\left(\p_y \sigma_i 
+ g_i^2\left(|H_i|^2 - v_i^2\right)\right)^2
+ \left|\D_y H_i + \sigma_i H_i - H_i M_i\right|^2 \non
&+& v_i^2 \p_y \sigma_i - \p_y
\left((\sigma_i H_i - H_i M_i)H_i^\dagger\right) \non
&\ge& v_i^2 \p_y \sigma_i - \p_y
\left((\sigma_i H_i - H_i M_i)H_i^\dagger\right).
\eeq
Thus the Hamiltonian is bounded from below. 
This bound is called Bogomol'nyi bound, and is saturated 
when the following BPS equations are satisfied 
\beq
\p_y \sigma_i + g_i^2\left( |H_i|^2 - v_i^2\right) = 0,
\quad
\D_y H_i + \sigma_i H_i - H_i M_i=0.
\label{eq:BPSeq_AH}
\eeq
In order to obtain the domain wall solution 
interpolating the two vacua in Eq.~(\ref{eq:vacua_AH}), 
we impose the boundary conditions :
\begin{eqnarray}
(H_{iL},H_{iR},\sigma_i) &=& \left(0,v_i,-m_i\right), \; \; y=-\infty, 
\nonumber \\
(H_{iL},H_{iR},\sigma_i) &=& \left(v_i,0,m_i\right), \; \quad y=\infty.
\label{eq:boundary_cond}
\end{eqnarray}
Tension $T_i$ of the domain wall is given by a topological 
charge as 
\beq
T_i = \int^\infty_{-\infty} dy\ \left[
v_i^2 \p_y \sigma_i - \p_y
\left((\sigma_i H_i - H_i M_i)H_i^\dagger\right)
\right]^\infty_{-\infty} = 2 m_i v_i^2.
\eeq

\begin{figure}[ht]
\begin{center}
\begin{tabular}{cc}
\includegraphics[height=4cm]{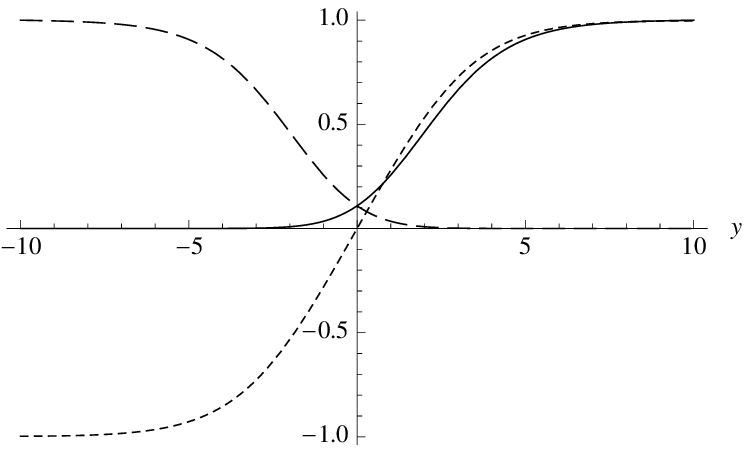} 
\includegraphics[height=4cm]{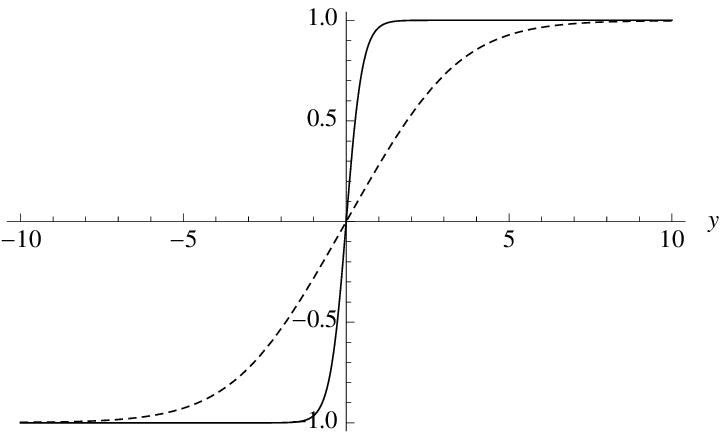}  
\end{tabular}
\caption{The left panel shows profiles of $H_{iL}$ (solid line), 
$H_{iR}$ (long-dashed line), 
and $\sigma_i$ (dashed-line) with finite 
gauge coupling ($g_i=0.5$). 
The right panel shows a plot of $\sigma_i$: 
dashed curve for finite ($g_i=0.5$) gauge coupling 
and solid curve for strong gauge coupling ($g_i=\infty$). 
The other parameters are $m_i=v_i=1$. }
\label{fig:dw_sample}
\end{center}
\end{figure}

The second equation of the BPS equations (\ref{eq:BPSeq_AH}) 
can be solved by the moduli matrix formalism  
\cite{Isozumi:2004jc, Eto:2006pg} 
with the constant matrix (vector) $H_{i0} = (C_{iL},C_{iR})$ 
\beq
H_i = v_i e^{-\frac{\psi_i}{2}} H_{i0}e^{M_iy},\quad
\sigma_i + i w_i = \frac{1}{2} \p_y \psi_i.
\label{eq:mm_AH}
\eeq
For a given $H_{i0}$, the scalar function $\psi_i$ is determined 
by the master equation
\beq
\p_y^2\psi_i = 2 g_i^2v_i^2 \left(1 - e^{-\psi_i} 
H_{i0} e^{2M_iy} H_{i0}^\dagger\right).
\label{eq:master_Abelian-Higgs}
\eeq
The asymptotic behavior of the field $\psi_i$ is determined 
by the condition that the configuration reaches 
the vacuum at left and right infinities:
\beq
\psi_i \to \log H_{i0} e^{2M_iy}H_{i0}^\dagger, 
\quad |y|\to\infty.
\eeq
There exists redundancy in the decomposition in 
Eq.~(\ref{eq:mm_AH}), which is called the $V$-transformation:
\beq
H_{i0} \to V_i H_{i0},\quad \psi_i \to \psi_i + 2 \log V_i,
\quad V_i \in \mathbb{C}^*.
\label{eq:V_Abelian-Higgs}
\eeq
For example, a single domain wall solution centered at $y=0$ 
can be generated by a moduli matrix
\beq
H_{i0} = (1,1).
\eeq
Then the master equation is 
\beq
\p_y^2 \psi_i = 2 g_i^2v_i^2 \left(1 - e^{-\psi_i}
\left(e^{2m_iy} + e^{-2m_iy}\right)\right).
\eeq
No analytic solutions for the master equation have been found 
for finite gauge couplings $g_i$, 
so we must solve it numerically.
The corresponding solution is shown in Fig.~\ref{fig:dw_sample}.
The generic solutions of the domain wall are generated by 
the generic moduli matrices (after fixing the $V$-transformation)
\beq
H_{i0} = \left(C_{iL}, C_{iR}\right),\quad C_{iL}, C_{iR} \in \mathbb{C}^*.
\eeq
The complex constants $C_{iL}, C_{iR}$ are free parameters 
containing the moduli parameters of the BPS solutions.
The moduli parameter can be defined by 
\beq
C_i \equiv \sqrt{\frac{C_{iR}}{C_{iL}}} = e^{i\alpha_i} e^{m_i y_i}. 
\eeq
The other degree of freedom in $C_{iL}, C_{iR}$ can be 
eliminated by the $V$-transformation in Eq.~(\ref{eq:V_Abelian-Higgs}) 
and has no physical meaning. 
Then the master equation is found to be
\beq
\p_y^2 \psi_i = 2 g_i^2 v_i^2 \left(1 - e^{-\psi_i}
\left(e^{2m_i(y-y_i)} + e^{-2m_i(y-y_i)}\right)\right).
\label{eq:master_Abelian-Higgs2}
\eeq
It is obvious that the real parameter $y_i$ is the 
translational moduli of the domain wall.
The other parameter $\alpha_i$ is an internal moduli 
which is the Nambu-Goldstone (NG) mode associated with
the $U(1)_{iA}$ flavor symmetry spontaneously broken 
by the domain walls.

One can take, if one wishes, the strong gauge coupling 
limit of the Lagrangian $\Lag_i$. 
As is well-known, the $U(1)$ gauge theory with two flavors of 
Higgs scalars in the strong gauge coupling limit 
becomes a non-linear sigma model whose target 
space is $\mathbb{C}P^1$:
\beq
|H_i|^2 = |H_{iL}|^2 + |H_{iR}|^2 = v_i^2.
\eeq
The gauge fields and the neutral scalar field become 
infinitely massive and lose their kinetic terms. 
They are mere Lagrange multipliers in the limit, and are solved as 
\beq
w_M^i = -\frac{i}{2v_i^2}\left(H_i\p_M H_i^\dagger 
- \p_MH_i H_i^\dagger\right),\quad
\sigma_i = \frac{1}{v_i^2} H_i M_i H_i^\dagger.
\eeq
Plugging these into $\Lag_i$, we get
\beq
\Lag_i^\infty = \p_MH_i P_i \p^M H_i^\dagger  
- H_iM_iP_iM_iH_i^\dagger,
\eeq
with a projection operator
\beq
P_i \equiv {\bf 1} - \frac{1}{v_i^2} H_i^\dagger H_i.
\eeq
Let us introduce an inhomogeneous coordinate $\phi_i$ 
of $\mathbb{C}P^1$ by
\beq
H_{iL} = \frac{v_i}{\sqrt{1+|\phi_i|^2}},\quad
H_{iR} = \frac{v_i \phi_i}{\sqrt{1+|\phi_i|^2}}.
\eeq
Then the Lagrangian of the $\mathbb{C}P^1$ model in terms 
of $\phi_i$ is
\beq
\Lag_i^\infty = v_i^2 \frac{|\p_M \phi_i|^2 - 4m_i^2 
|\phi_i|^2}{\left(1+|\phi_i|^2\right)^2} .
\eeq
Let us reconsider the domain wall solutions in this limit. 
The Hamiltonian can be written as 
\beq
{\cal  H}_i^\infty 
&=& \frac{v_i^2}{(1+|\phi_i|^2)^2} \left| \p_y \phi + 2m_i  
\phi_i\right|^2
+ 2m_i v_i^2\frac{d}{dy} \frac{1}{1+|\phi_i|^2} \non
&\ge& 2m_i v_i^2\frac{d}{dy} \frac{1}{1+|\phi_i|^2}.
\eeq

The BPS equation and the boundary conditions are given by 
\beq
\p_y \phi_i + 2m_i \phi_i &=& 0,
\nonumber \\
\phi(y=-\infty)=\infty, &\quad & \phi(y=\infty)=0 ,
\eeq
corresponding to the boundary conditions in 
Eq.(\ref{eq:boundary_cond}). 
The BPS equation can be easily solved by
\beq
\phi_i = C_{iL}^{-1}C_{iR} e^{-2m_iy} = C_i^2 e^{-2m_iy}.
\label{eq:sol_u1}
\eeq
The tension of the domain wall is
\beq
T_i = \int^\infty_{-\infty} dy\ {\cal H}_i^\infty = 2m_i v_i^2.
\eeq
This is the same as the one in the finite gauge coupling model.

In this way, the strong gauge coupling limit has a great 
advantage compared to the finite gauge coupling case. 
One can exactly solve the BPS equation and see the moduli 
parameter in the analytic solutions.
Furthermore, there is no important differences between 
domain wall solutions in the finite coupling (Abelian-Higgs model) 
and the strong coupling (non-linear sigma model).
 Both solutions have the same tension of domain wall 
and the same number of the moduli parameters.
To see the difference explicitly, let us compare the 
configuration of the neutral scalar field $\sigma_i$.
In the strong gauge coupling limit, it can be written as 
\beq
\sigma_i = m_i \frac{1-|\phi_i|^2}{1+|\phi_i|^2} 
= m_i \tanh 2m_i(y-y_i),
\eeq
where we have used
\beq
C_i = e^{i\alpha} e^{m_iy_i}.
\label{eq:sol_sample}
\eeq
In Fig.~\ref{fig:dw_sample}, we show the configurations of 
$\sigma_i$ in two cases, the one in the small finite 
gauge coupling and the one in the strong gauge coupling limit. 
As can be seen from the figure, there are no significant 
differences.

Let us next derive the low energy effective theory on 
the domain wall.
We integrate all the massive modes while keeping the 
massless modes. 
We use the so-called moduli approximation where the 
dependence on (3+1)-dimensional spacetime coordinates 
comes into the 
effective Lagrangian only through  the moduli fields:
\beq
C_i \to C_i(x^\mu),\quad
\phi_i(y) \to \phi_i(y,C_i(x^\mu)) = C_i(x^\mu)^2e^{-2m_iy}.
\eeq
The effective Lagrangian for the moduli field $C_i(x^\mu)$ 
can be obtained by plugging this into the Lagrangian $\Lag_i$ 
and integrate it over $y$. 
This can be done explicitly as follows.
\beq
\Lag_{i,{\rm eff}} =  \int^\infty_{-\infty} dy 
\frac{v_i^2}{\left(|C_i|^{-2}e^{2m_iy}+|C_i|^2 e^{-2m_iy}\right)^2} 
\frac{\left|\p_\mu C_i^2\right|^2}{|C_i^2|^2}
= \frac{v_i^2}{4m_i } 
\frac{\left|\p_\mu C_i^2\right|^2}{\left|C_i^2\right|^2}.
\label{eq:eff_Abelian-Higgs0}
\eeq
With Eq.~(\ref{eq:sol_sample}), the effective Lagrangian is given by
\beq
\Lag_{i,{\rm eff}} 
= \frac{2m_iv_i^2}{2}(\p_\mu y_i)^2 
+ \frac{v_i^2}{m_i}(\p_\mu \alpha_i)^2,
\label{eq:effLagAbelian}
\eeq
where energy of soliton solution is neglected since it does not contribute
to dynamics of moduli.
Note that $2m_iv_i^2$ is precisely the domain wall tension.
This is the free field Lagrangian.

Although we have derived this effective Lagrangian in the 
strong gauge coupling limit, we can obtain the same 
Lagrangian in the finite gauge coupling constant. 
In other words, the effective  Lagrangian cannot 
distinguish the infinite versus finite coupling cases 
at least in the quadratic order of the derivative expansion.

\smallskip
\subsection{Localization of the Abelian gauge fields 
\label{sec:localization_Abelian}}

In the previous subsection, we have seen 
that the NG modes of the translation and $U(1)$ global 
symmetry are the only massless modes in the Abelian-Higgs 
model. 
They are localized on the domain wall. 
There are no massless gauge field on the domain wall and 
all the modes contained in the gauge field are massive. 
The mass of the lightest mode of the gauge field is of the 
order of the inverse of the width of the domain wall, 
since the bulk outside of the domain wall is in the Higgs 
phase. 
The low energy effective Lagrangian for the massless fields 
is obtained after integrating out the massive modes including 
gauge fields. 

In order to obtain the massless gauge field to be localized 
on the domain wall, we need a new gauge symmetry which is 
unbroken in the bulk.
Recently, a new mechanism was proposed to localize 
gauge fields on domain walls \cite{OhSa}.

A key ingredient is the so-called dielectric coupling 
constant \cite{Kogut:1974sn,Fukuda:1977wj} for the new 
gauge symmetry. 
To illustrate the new localization mechanism, let us 
introduce a new $U(1)$ gauge field $a_M$ which we wish to 
localize on the domain wall. 
Since this gauge symmetry should be unbroken in the bulk, 
we consider the case where all the Higgs fields are neutral 
under this newly introduced $U(1)$ gauge symmetry. 
The gauge field $a_M$ is assumed to couple to the neutral 
scalar fields $\sigma_i$ only in the following particular 
combination 
\beq
\Lag = \Lag_1 + \Lag_2 - \frac{\lambda}{2}
\left(\frac{\sigma_1}{m_1} 
- \frac{\sigma_2}{m_2}\right)\left({\cal G}_{MN}\right)^2,
\label{eq:lag_Abelian-Higgs_local}
\eeq
where a real constant $\lambda$ with the unit mass 
dimension, in accordance with the (4+1)-dimensional 
spacetime and the field strength is defined by 
\begin{equation}
{\cal G}_{MN} = \p_M a_N - \p_N a_M. 
\end{equation}
The field-dependent gauge coupling function is given by
\beq
\frac{1}{4e^2(\sigma)} = \frac{\lambda}{2}
\left(\frac{\sigma_1}{m_1} - \frac{\sigma_2}{m_2}\right), 
\label{eq:coupling_Abelian-Higgs}
\eeq
which depends on the position $y$ through fields $\sigma_i$.  
Thus the field-dependent gauge coupling function 
$e(\sigma)$ plays the role 
of the dielectric coupling constant.
Furthermore, the special choice in Eq.~(\ref{eq:coupling_Abelian-Higgs}) 
is chosen for the gauge interaction to 
become strongly coupled in the bulk ($\sigma_i \to \pm m_i$ 
as $y \to \pm \infty$).

Let us again consider a double copy of domain walls as a 
background configuration 
in the Abelian-Higgs model in Eq.~(\ref{eq:lag_Abelian-Higgs_local}). 
Since Lagrangian has no term 
linear in $a_M$, the equations of motion for 
$a_M$ is trivially solved by $a_M=0$, and the rest of 
the equations of motion are explicitly the same as 
those in the previous subsection. 
Therefore the domain wall solution in the previous 
subsection together with $a_M=0$ is still a solution 
of the equations of motion. 
Clearly, the low energy effective Lagrangian on the 
domain wall is also unchanged 
\beq
\Lag_{\rm eff} = \Lag_{1,{\rm eff}} + \Lag_{2,{\rm eff}} 
- \frac{1}{4e_4^2} ({\cal G}_{\mu\nu})^2,
\eeq
except for the additional kinetic term 
(the last term) of the (3+1)-dimensional gauge 
field $w_{\mu}$, which is the zero mode 
($y$-independent mode) of 
the (4+1)-dimensional field $w_{\mu}$. 
The (3+1)-dimensional gauge coupling constant is given by 
\beq
 \frac{1}{4e_4^2} = \frac{\lambda}{2}\int^\infty_{-\infty} 
dy\ \Bigl(\frac{\sigma_1}{m_1} - \frac{\sigma_2}{m_2}\Bigr)
 = \frac{\lambda}{4}\left[\frac{\psi_1}{m_1} 
- \frac{\psi_2}{m_2}\right]^\infty_{-\infty} 
= \lambda(y_2-y_1),
\label{eq:integral4Dcoupling}
 \eeq
 where we 
have used the asymptotic behavior 
$\psi_i \to \log 2\cosh 2m_i(y-y_i)$ as 
$|y|\to \infty$.
 Note that this result is again independent of the gauge 
couplings $g_i$ in the domain wall sector.
 In summary, the low energy effective Lagrangian is
 \beq
 \Lag_{\rm eff} = 
 \sum_{i=1,2}\left[\frac{2m_iv^2}{2}(\p_\mu y_i)^2 
+ \frac{v_i^2}{m_i}
(\p_\mu \alpha_i)^2\right] 
 - \lambda (y_2-y_1) ({\cal G}_{\mu\nu})^2.
 \eeq
Now we separate the quantum fields (fluctuations) from the 
classical background moduli parameters by
\beq
y_i(x^\mu) = y_i^0 + \delta y_i,\quad
\alpha_i(x^\mu) = \alpha_i^0 + \delta \alpha_i.
\label{eq:separate_Abelian-Higgs}
\eeq
Then the effective Lagrangian 
up to the second order of the small quantum fluctuations 
is given by
 \beq
 \Lag_{\rm eff}(y_i^0,\alpha_i^0) = 
 \sum_{i=1,2}\left[\frac{2m_iv_i^2}{2}(\p_\mu \delta y_i)^2 
+ \frac{v_i^2}{m_i}(\p_\mu \delta \alpha_i)^2\right] 
 - \lambda(y_2^0-y_1^0) ({\cal G}_{\mu\nu})^2 .
 \label{eq:eff_Abelian-Higgs}
 \eeq
We note that the massless gauge field $a_\mu$ has a 
positive finite gauge coupling squared\footnote{ 
Here we are content with the fact that the positivity of 
the gauge kinetic term is assured at least in finite region 
of moduli space, instead of just at a point. 
However, it is possible to make a more economical model 
where one has less moduli, and the positivity of the 
gauge kinetic term is assured\cite{OhSa}. 
} 
$1/(4\lambda(y_2^0-y_1^0))$ provided $y_2^0-y_1^0>0$. 

Although we succeeded in localizing the massless $U(1)$ 
gauge field $a_\mu$ on the domain walls, the Lagrangian 
Eq.~(\ref{eq:eff_Abelian-Higgs}) has no charged matter fields 
minimally coupled with the localized gauge field $a_\mu$. 
To obtain matter fields interacting with the localized 
gauge field, one may be tempted to identify the Higgs 
fields $H_i=(H_{iL}, H_{iR})$ as matter fields \footnote{
We consider the diagonal subgroup $U(1)_A$ of $U(1)_{1A}$ 
and $U(1)_{2A}$. 
Actually the $U(1)_{iA}$ global symmetries are broken 
by the domain wall solution, we consider this gauging 
to leading order of gauge coupling only to illustrate 
the Higgs mechanism for the broken symmetry. 
} with charges $(1, -1)$. 
The minimal gauge interaction of Higgs 
fields with the $a_M$ is introduced through 
the modified covariant derivatives as 
\beq
\tilde{\D}_M H_{iL} &=& \p_M H_{iL} + i w_M^i H_{iL} 
+ i a_M H_{iL},\\
\tilde{\D}_M H_{iR} &=& \p_M H_{iR} + i w_M^i H_{iR} 
- i a_M H_{iR}. 
\eeq
Since the moduli field $C_i$ is charged, the 
derivatives in the low energy effective 
theory Eq.~(\ref{eq:eff_Abelian-Higgs0}) should be replaced by the 
covariant derivative 
\beq
\p_\mu C_i \to \D_\mu C_i = \p_\mu C_i + i a_\mu C_i.
\eeq
It is straightforward task to derive
the effective Lagrangian with the covariant derivative above
along the same line of reasoning for the previous case
 \beq
 \Lag_{\rm eff}(y_i^0,\alpha_i^0) & = &
 \sum_{i=1,2}\left[\frac{2m_iv_i^2}{2}(\p_\mu \delta y_i)^2 
+ \frac{v_i^2}{m_i}(\p_\mu \delta \alpha_i + q_i a_\mu)^2\right] 
\nonumber \\
 &-& \lambda(y_2^0-y_1^0) ({\cal G}_{\mu\nu})^2. 
 \eeq
This clearly shows that 
the new gauge field $a_\mu$ is 
not massless due to the Higgs mechanism, 
and should be integrated out together with 
the other massive fields. 
Namely the low energy effective Lagrangian 
does not include the massless gauge fields, since 
the $U(1)$ symmetry which we gauged is broken by the 
domain wall.
A more explicit example at the strong gauge coupling limit 
is described in Appendix \ref{app3}.
Thus the Abelian-Higgs model in this section gives an 
important lesson that we should not gauge a symmetry which 
is broken by the domain wall solution, since the 
corresponding gauge fields may be localized on the domain 
walls but they become massive and should be integrated out 
from the low energy effective theory. 
In the next section, we will give a model with 
a non-Abelian global symmetry whose unbroken subgroup 
can be gauged to yield massless localized gauge fields 
on the domain wall.

\bigskip
\section{The chiral model 
\label{sec:chiral model}}

In this section we study domain walls in the chiral model 
which is a natural extension of the Abelian-Higgs model 
in the previous section. 
This chiral model leads to two important consequences 
1) massless non-Abelian gauge fields are localized on the 
domain wall and moreover 
2) the scalar fields which are non-trivially interacting 
are also localized on the domain walls. 

\smallskip
\subsection{The domain walls in the chiral model}

As a natural extension of the domain wall sector in the 
previous section, we consider the Yang-Mills-Higgs model 
with $SU(N)_{c}\times U(1)$ gauge symmetry with 
$S[U(N)_{L}\times U(N)_{R}] = SU(N)_{L}\times SU(N)_{R} 
\times U(1)_{A}$ flavor symmetry \cite{ShYu2,EtFuNiOhSa}. 
To localize the gauge field in a simple manner, 
we again introduce two sectors $\Lag_1$ and $\Lag_2$, but 
only the former is extended to 
Yang-Mills-Higgs system and the latter is the same form 
as (\ref{Abelian-Higgs}).
The second sector couples to the first sector through 
the coupling as described in 
(\ref{eq:lag_Abelian-Higgs_local}) after gauging the flavor symmetry 
 it plays a role as localization of gauge fields, combined 
with the first sector.
The matter contents are summarized in Table \ref{table:SY}. 
Since the presence of two factors of $SU(N)$ global 
symmetry resembles the chiral symmetry of QCD, we call 
this Yang-Mills-Higgs system as the chiral model. 
\begin{table}
\begin{center}
\begin{tabular}{c|cccccccc}
\hline
  & $SU(N)_{c}$ & $U(1)_1$ & $U(1)_2$ & $SU(N)_{L}$ 
& $SU(N)_{R}$ & $U(1)_{1A}$ & $U(1)_{2A} $ & mass\\ \hline
$H_{1L}$ & $\square$ & 1 & 0 & $\square$ & {\bf 1} 
& 1 & 0 & $m_1{\bf 1}_{N}$\\ 
$H_{1R}$ & $\square$ &  1 & 0 & {\bf 1} 
& $\square$ & $-1$ & 0 & $- m_1{\bf 1}_{N}$ \\ 
$\Sigma_1$ & ${\rm adj}\oplus{\bf 1}$ & 0 & 0 & {\bf 1} 
& {\bf 1} & 0 & 0 & 0\\
$H_{2L}$ & {\bf 1} & 0 & 1 & {\bf 1} 
& {\bf 1} & 0 & $1$  & $m_2$\\ 
$H_{2R}$ & {\bf 1} & 0 & 1 & {\bf 1} 
& {\bf 1}& 0 & $-1$ & $- m_2$ \\ 
$\Sigma_2$ & {\bf 1} & 0 & 0 & {\bf 1} 
& {\bf 1} & 0 & 0 & 0\\
\hline
\end{tabular}
\end{center}
\caption{Quantum numbers of the domain wall sectors in the 
chiral model. 
}
\label{table:SY}
\end{table}

The Lagrangian is then given by
\beq
\Lag &=& \Lag_1+\Lag_2, 
\label{eq:Lag_chi_sum}\\
\Lag_1 &=& \Tr\left[-\frac{1}{2g_1^2}(F_{1MN})^2 
+ \frac{1}{g_1^2}(\D_M\Sigma_1)^2 
+ \left|\D_M H_1\right|^2 \right] - V_1,
\label{eq:Lag_chi}\\
V_1 &=& \Tr\left[ \frac{g_1^2}{4} \left(H_1 H_1^\dagger 
- v_1^2 {\bf 1}_{N}\right)^2 + \left|\Sigma_1 H_1
-H_1M_1\right|^2\right],
\label{eq:YM_Higgs_Lag}
\eeq
with $H_1 = \left(H_{1L},\ H_{2L}\right)$.
$\Lag_2$ is the same form as (\ref{Abelian-Higgs}) with $i=2$.
Gauge fields of $U(N)_c=(SU(N)_c\times U(1)_1)/Z_{\bf N}$ 
are denoted as $W_{1M}$, and adjoint scalar as 
$\Sigma_{1}$. 
The covariant derivative and the field strength are denoted as 
$\D_M\Sigma_1 = \p_M \Sigma_1 + i \left[W_{1M}, \Sigma_1\right]$, 
$\D_M H_1 = \p_M H_1 + i W_{1M} H_1$, 
and $F_{1MN} = \p_M W_{1N} - \p_N W_{1M} 
+ i \left[W_{1M},W_{1N}\right]$. 
The mass matrix is given by 
$M_1 = {\rm diag}\left(m_1 {\bf 1}_N, -m_1{\bf 1}_N\right)$. 
Let us note that the chiral model reduces to the Abelian-Higgs 
model in the limit of $N\to1$, by deleting all the 
$SU(N)$ groups. 

The second sector is just necessary to realize the 
field-dependent gauge coupling function similar to 
(\ref{eq:lag_Abelian-Higgs_local}) as we will discuss in the 
subsequent subsection. 
In the rest of this subsection, we focus only on the 
first sector ($i=1$) and suppress the index $i=1$.
The symmetry transformations act on the fields as
\beq
H = \left(H_{L},H_{R}\right) &\to& U_{c}\left(H_{L},H_{R}\right)
\left(
\begin{array}{cc}
U_{L}e^{i\alpha} & \\
& U_{R}e^{-i\alpha}
\end{array}
\right),
\label{eq:transf_H}
\\
\Sigma &\to& U_{c}\Sigma U_{c}^\dagger,
\label{eq:transf_sigma}
\eeq
with $U_{c} \in U(N)_{c}$, $U_{L} \in SU(N)_{L}$, 
$U(N)_{R} \in SU(N)_{R}$ and $e^{i\alpha} \in U(1)_{A}$.

There exist $N+1$ vacua in which the fields develop the 
following VEV
\beq
H &=& (H_{L}, H_{R}) = v\left(
\begin{array}{cc|cc}
{\bf 1}_{N-r} & & {\bf 0}_{N-r} & \\
& {\bf 0}_{r} & & {\bf 1}_{r}
\end{array}
\right),\\
\Sigma&=& m \left(
\begin{array}{cc}
{\bf 1}_{N-r} & \\
& -{\bf 1}_{r}
\end{array}
\right),
\eeq
with $r=0,1,2,\cdots, N$.
We refer these vacua with the label $r$. 
In the $r$-th vacuum, both the local gauge symmetry $U(N)_c$ 
and the global symmetry are broken, but a diagonal global 
symmetries are unbroken (color-flavor-locking) 
\begin{eqnarray}
&&U(N)_c \times SU(N)_L \times SU(N)_R \times U(1)_{A} 
\to 
\nonumber \\
&&SU(N-r)_{L+c}\times SU(r)_L \times 
SU(r)_{R+c}\times SU(N-r)_R\times U(1)_{A+c}.
\label{eq:symmetry_r_vacuum}
\end{eqnarray}

As in the Abelian-Higgs model, the BPS equations for the 
domain walls can be obtained through the 
Bogomol'nyi completion of the energy density with the 
assumption that all the fields depend on only the
fifth coordinate $y$ and $W_{\mu} = 0$:
\beq
{\cal H} &=& \Tr\left[
\frac{1}{g^2}\left(\D_y\Sigma 
- \frac{g^2}{2}\left(v^2 {\bf 1}_N 
- HH^\dagger\right)\right)^2
+\left|\D_yH+\Sigma H-HM\right|^2
\right]\non
&+& \p_y \left\{ \Tr \left[v^2 \Sigma 
- \left(\Sigma H-HM\right)H^\dagger\right]\right\} \non
&\ge&  \p_y \left\{ \Tr \left[v^2 \Sigma 
- \left(\Sigma H-H M \right)H^\dagger\right]\right\}.
\eeq
This bound is saturated when the following BPS equations 
are satisfied 
\beq
\D_y\Sigma - \frac{g^2}{2}\left(v^2 {\bf 1}_{N} 
- H H^\dagger\right) = 0,\label{eq:BPS_SY1}\\
\D_y H+\Sigma H-H M = 0.\label{eq:BPS_SY2}
\eeq
The tension of the domain wall is given by 
\beq
T &=& \int^\infty_{-\infty} dy \ \p_y 
\left\{\Tr \left[v^2\Sigma 
- \left(\Sigma H-H M\right)H^\dagger\right]\right\} \non
&=& v^2 \; {\rm Tr} \left[\Sigma(+\infty) 
- \Sigma(-\infty)\right].
\label{eq:tension_SY}
\eeq 

Let us concentrate on the domain wall which connects the 
0-th vacuum at $y\to \infty$ and the $N$-th vacuum at 
$y \to -\infty$. Its tension can be read as 
\beq
T = 2N v^2m,
\eeq
from
Eq.~(\ref{eq:tension_SY}). 
Since there are $N+1$ possible vacua, the maximal number of 
walls is $N$ at various positions. 
The simplest domain wall solution corresponding to the 
coincident walls is given by making an 
ansatz that $H_{L}$, $H_{R}$, $\Sigma$ and $W_{y}$ 
are all proportional to the unit matrix. 
Then the BPS equations (\ref{eq:BPS_SY1}) and 
(\ref{eq:BPS_SY2}) can be identified with 
the BPS equations in Eq.~(\ref{eq:BPSeq_AH}) in  the 
Abelian-Higgs model.
Thus the domain wall solution can be solved as
\begin{align}
H_{L} &= v e^{-\frac{\psi}{2}}e^{m y}~{\bf 1}_{N},\label{eq:sol1_SY}\\
H_{R} &= v e^{-\frac{\psi}{2}}e^{-m y}~{\bf 1}_{N},\label{eq:sol2_SY}\\
\Sigma + i W_{y} &= \frac{1}{2}\p_y\psi{\bf 1}_{N}\label{eq:sol3_SY},
\end{align}
where $\psi$ is the solution of the master equation 
(\ref{eq:master_Abelian-Higgs}) in the Abelian-Higgs model.
Eq.(\ref{eq:symmetry_r_vacuum}) shows that the unbroken 
global symmetry for $N$-th vacuum ($H_{L} = 0$, 
$H_{R} = v{\bf 1}_{N}$ and $\Sigma = -m{\bf 1}_{N}$) 
at the left infinity $y \to -\infty$ is 
$SU(N)_L \times SU(N)_{R+c} \times U(1)_{A+c}$, 
whereas that for the $0$-th vacuum ($H_{L} = v{\bf 1}_{N}$, 
$H_{R} = 0$ and $\Sigma = m{\bf 1}_{N}$) 
at the right infinity $y\to\infty$ is 
$SU(N)_{L+c} \times SU(N)_R \times U(1)_{A+c}$.

The domain wall solution further breaks these unbroken 
symmetries because it interpolates the two vacua.
The breaking pattern by the domain wall is \footnote{
The unbroken generators of $U(1)_{A+c}$ for $r$-th 
vacuum contains different combination of 
$U(N)_c$ generators depending on $r$. 
Therefore the right and left vacua preserve actually 
different $U(1)_{A+c}$, and the wall solution does not 
preserve any of these $U(1)_{A+c}$. 
}
\beq
U(N)_c \times SU(N)_L \times SU(N)_R \times U(1)_{A} 
\to SU(N)_{L + R+c}.
\eeq
This spontaneous breaking of the global symmetry gives 
NG modes on the domain wall as massless degrees of freedom 
valued on the coset similarly to the chiral symmetry breaking 
in QCD : 
\beq
\frac{SU(N)_{L} \times SU(N)_R}{SU(N)_{L + R+c}} 
\times U(1)_{A}.
\label{eq:NGmodes_wall}
\eeq
Since our model can be embedded into a supersymmetric 
field theory, these NG modes ($U(N)$ chiral fields) appear as complex scalar fields
accompanied with additional $N^2$ pseudo-NG modes\footnote{
One of them is actually a genuine NG mode corresponding to 
the broken translation. }.

\subsection{Localization of the matter fields}
\label{sec:chiral}

In the remainder of this subsection, we will give 
the low-energy effective Lagrangian 
on the domain walls where the massless moduli fields (the matter fields) are localized.
The best way to parametrize these massless moduli 
fields is to use the moduli matrix 
formalism \cite{EtFuNiOhSa, Isozumi:2004jc, Eto:2006pg} 
\begin{align}
H_{L} & = v e^{my}S^{-1}\,,\label{eq:solU1_SY}\\
H_{R} & = v e^{-my} S^{-1}e^{\phi}\,,\label{eq:solU2_SY}\\
\Sigma + i W_{y} & = S^{-1}\partial_y S\,,
\label{eq:solU3_SY}
\end{align}
where $S\in GL(N,{\bf C})$ and $\Omega=S S^\dagger$ is the 
solution of the following 
master equation 
\begin{eqnarray}
\partial_y\left( \Omega^{-1}\partial_y \Omega\right) = g^2v^2 
  \left(\mathbf{1}_N - \Omega^{-1}\Omega_0\right)\,, 
\label{eq:wll:master-eq-wall}
\end{eqnarray}
where
\begin{equation*}
\Omega_0 
= e^{2my}\mathbf{1}_N+e^{-2my}e^{\phi}e^{\phi^{\dagger}}\,.
\label{eq:omega0}
\end{equation*}
We have used the $V$-transformation to identify 
the moduli $e^\phi$, which is a complex $N$ by $N$ matrix. 
It can be parametrized by an $N \times N$ hermitian matrix 
$\hat x$ 
and a unitary matrix $U$ as \cite{EtFuNiOhSa} 
\begin{eqnarray}
e^{\phi} = e^{\hat x}U^{\dagger}, 
\label{eq:xu_decomp}
\end{eqnarray}
where $U$ is nothing but the $U(N)$ chiral fields associated with
the spontaneous symmetry breaking Eq.~(\ref{eq:NGmodes_wall}) and
$\hat x$ is the pseudo-NG modes whose existence we promised above.

In the strong gauge coupling limit $g \to \infty$, 
solution of master equation is simply $\Omega = \Omega_0$. 
After fixing the $U(N)_c$ gauge, 
we obtain
\begin{eqnarray}
S=e^{\hat x/2}\sqrt{2\cosh (2my-\hat x)}\,. 
\end{eqnarray} 
Let us denote, for brevity 
\begin{equation}
\hat y= 2my-\hat x\,, 
\end{equation}
the Higgs fields are then given as 
\begin{align}
H_{L} & = v \frac{e^{\hat y/2}}{\sqrt{2\cosh \hat y}}\,,
\label{eq:solU1_SY3}\\
H_{R} & = v \frac{e^{-\hat y/2}}{\sqrt{2\cosh \hat y}} 
U^\dagger\,.
\label{eq:solU2_SY3}
\end{align}
From this solution, one can easily recognize that eigenvalues of $\hat x$
correspond to the positions of the $N$ domain walls in the $y$ direction.
Now we  promote moduli parameters $\hat x$ and 
$U$ to fields on the domain wall world volume, namely 
functions of world volume coordinates $x^\mu$. 
We plug the domain wall solutions 
$H_{L,R}(y; \hat x(x^\mu),U(x^\mu))$ 
into the original Lagrangian $\Lag$ in 
Eq.(\ref{eq:Lag_chi}) at $g \to \infty$ 
and pick up the terms quadratic in the derivatives. 
Thus the low energy effective Lagrangian is given by \cite{EtIsNiOhSa}
\beq
\Lag_{{\rm eff}} = \int_{-\infty}^{\infty} dy\,\Tr\left[
\partial_\mu H_{L}\partial^\mu H_{L}^\dagger 
+ \partial_\mu H_{R}\partial^\mu H_{R}^\dagger
-v^2 W_\mu W^\mu
\right]\,,
\eeq
where
\beq
 W_\mu={i \over 2v^2}\left[\partial_\mu H_LH_L^\dagger 
-H_L\partial_\mu H_L^\dagger+(L\leftrightarrow R)\right].
\eeq
Here we have eliminated the massive gauge field $W_\mu$ 
by using the equation of motion.
Using the solutions for $H_L$ and $H_R$ 
we have found a closed formula for the effective Lagrangian 
up to the second order of derivatives but with full nonlinear 
interactions involving moduli fields $\hat x$ and $U$. 
Detailed derivation is given in Appendix \ref{app2}.

Here we exhibit the result only in the leading orders of 
$U-1$ and $\hat x$:
\begin{equation}
\oper{L}_{\rm eff} = \frac{v^2}{2m}\Tr
\Bigl(\partial_{\mu}U^{\dagger}\partial^{\mu}U
+\partial_{\mu}\hat x\partial^{\mu}\hat x\Bigr)+\ldots
\end{equation}  
When $N = 1$ and with the redefinitions $U = e^{2i\alpha_1}$, 
and $\hat x = 2m y_1$, this coincides with the effective 
Lagrangian ${\cal L}_{i=1,{\rm eff}}$ in 
Eq.(\ref{eq:effLagAbelian}) of the Abelian-Higgs model, 
which we obtained in the previous section.

\smallskip
\subsection{Localization of the gauge fields}

Let us next introduce the gauge fields which are to be 
massless and localized on the domain walls. 
As we learned in section~\ref{sec:Abelian-Higgs}, the associated gauge 
symmetry should not be broken by the domain walls. 
Therefore, the symmetry which we can gauge is the unbroken 
symmetry $SU(N)_{L+R+c}$ itself or its subgroup. 

Let us gauge $SU(N)_{L+R}\equiv SU(N)_V$ and let $A_{\mu}^a$ be the $SU(N)_{L+R}$ gauge field.
The  Higgs fields are in the bi-fundamental representation of $U(N)_c$ and $SU(N)_{L+R}$.
The covariant derivatives of the Higgs fields are modified by
\beq
\tilde\D_M H_{1L} = \p_M H_{1L} + i W_{1M} H_{1L} 
- i H_{1L}A_{M},\label{eq:covd1}\\
\tilde\D_M H_{1R} = \p_M H_{1R} + i W_{1M} H_{1R} 
- i H_{1R}A_{M}\label{eq:covd2}.
\eeq
The quantum numbers are summarized in Table \ref{table:gaugedSY}.
\begin{table}
\begin{center}
\begin{tabular}{c|ccccccc}
\hline
  & $SU(N)_{c}$& $U(1)_1$ & $U(1)_2$ &$SU(N)_{V}$  & $U(1)_{1A}$ & $U(1)_{2A}$ & mass\\ \hline
$H_{1L}$ & $\square$ & 1 & 0 & $\square$ & 1  & 0 & $m_1{\bf 1}_{N}$\\ 
$H_{1R}$ & $\square$ & 1 & 0 & $\square$ & $-1$ & 0 & $- m_1{\bf 1}_{N}$ \\ 
$\Sigma_1 $ & ${\rm adj}\oplus{\bf 1}$  & 0 & 0 & {\bf 1} & 0 & 0 & 0 \\
$H_{2L}$ & {\bf 1} & 0 & 1 & {\bf 1} & 0 & 1  & $m_2$\\ 
$H_{2R}$ & {\bf 1} & 0 & 1 & {\bf 1} & 0 & $-1$ & $- m_2$ \\ 
$\Sigma_2 $ & {\bf 1} & 0 & 0 & {\bf 1} & 0 & 0 & 0  \\

\hline
\end{tabular}
\end{center}
\caption{Quantum numbers of the domain wall sectors in gauged chiral model}
\label{table:gaugedSY}
\end{table}

We now introduce a field-dependent gauge coupling function 
$g^2(\Sigma)$ for $A_{M}$, which is inspired by 
the supersymmetric model in Ref.\citen{OhSa}. 
\beq
\frac{1}{2e^2(\Sigma)} = \frac{\lambda}{2}
\left(\frac{\Tr\Sigma_1}{Nm_1} 
- \frac{\Sigma_2}{m_2}\right).
\label{Lag_gChi2}
\eeq
The Lagrangian is given by 
\beq
\Lag = \tilde{\Lag_1}+\Lag_2
 - \frac{1}{2e^2(\Sigma)}
\Tr\left[G_{MN}G^{MN}\right]. 
\label{Lag_gChi}
\eeq
The $\tilde \Lag_1$ in Eq.~(\ref{Lag_gChi}) 
is given by Eq.~(\ref{eq:Lag_chi}) where the covariant 
derivatives are replaced with those in Eqs.~(\ref{eq:covd1}) 
and (\ref{eq:covd2}).

We first wish to find the domain wall solutions in this 
extended model.
As before, we make ansatz that all the fields depend on 
only $y$ and $W_{\mu} = A_{\mu} =0$.
Let us first look at the equation of motion of the new 
gauge field $A_M$. It is of the form
\beq
\D_MG^{MN} = J^N,
\label{eq:eom_A}
\eeq
where $J_{M}$ stands for the current of $A_{M}$.
Note that the current $J_{M}$ is zero, by definition, if 
we plug the domain wall solutions
in the chiral model before gauging the $SU(N)_{L+R}$. 
This is because the domain wall 
configurations do not break $SU(N)_{L+R}$.
Therefore, $A_{M}=0$ is a solution of Eq.~(\ref{eq:eom_A}).

Then, we are left with equation of motion with $A_{M}=0$ 
which are identical to those in the ungauged chiral model 
in the previous subsections. 
Therefore the gauged chiral model admits 
the same domain wall solutions 
as those (Eqs.~(\ref{eq:solU1_SY3}) and (\ref{eq:solU2_SY3})) in the ungauged chiral model.

The next step is to derive the low energy effective theory on 
the domain wall world-volume 
in the moduli approximation as in the previous subsections.
Again, we promote the moduli parameters as the fields on the 
domain wall world-volume 
and pick up the terms up to the quadratic order of the 
derivative $\p_\mu$.
Similarly to section~\ref{sec:chiral}, we utilize 
the strong gauge coupling limit $g_i\to \infty$, 
to simplify the computation without changing the final 
result. 
Let us emphasize that we keep the field-dependent 
gauge coupling function $e(\Sigma)$ finite. 
The spectrum of massless NG modes is unchanged 
by switching on the $SU(N)_{L+R}$ gauge interactions 
\footnote{
Tree level mass spectra are unchanged even though the chiral 
symmetry $SU(N)_L \times SU(N)_R$ is 
broken by the $SU(N)_{L+R}$ gauge interactions. 
}. 

We just repeat the similar computation to those in section~\ref{sec:chiral}. 
Again we shall focus on the first sector ${\cal L}_1$ and suppress the index $i=1$ of fields.
Since color gauge fields $W_\mu$ becomes auxiliary fields 
and eliminated through their equations of motion, it is 
convenient to define the covariant derivative only for the 
flavor ($SU(N)_{L+R}$) gauge interactions as 
\beq
\hat D_\mu H = \p_\mu H - i H A_{\mu}.
\eeq
Then we obtain the effective Lagrangian of the first sector as
\begin{eqnarray}
\oper{L}_{1,\rm eff} &=& \int_{-\infty}^{\infty} dy \,
\mathrm{Tr}\Bigl[\hat D_{\mu}H_{L}(\hat D^{\mu}H_{L})^{\dagger}
+\hat D_{\mu}H_{R}(\hat D^{\mu}H_{R})^{\dagger}
-v^2W_{\mu}W^{\mu} \nonumber \\
&&-\frac{1}{2e^2(\Sigma)} G_{MN}G^{MN}\Bigr]\,,
\end{eqnarray}
with
\beq
 W_{\mu}={i \over 2v^2}\left[\hat{D}_\mu H_{L}H_{L}^\dagger
-H_{L}(\hat{D}_\mu H_{L})^\dagger+(L\leftrightarrow R)\right].
\eeq
Eliminating $W_\mu$, we obtain 
the following expression for the integrand of the 
effective Lagrangian after some simplification 
\begin{equation}\label{eq:lagr1}
\oper{L}_{\rm eff} = \frac{1}{2v^2}\int_{-\infty}^{\infty} 
dy\, \Tr\Bigl[\oper{D}_{\mu}H_{ab}\oper{D}^{\mu}H_{ba}
\Bigr]\,,
\end{equation}
where we defined fields $H_{ab}$ with the label $ab$ of adjoint 
representation of the flavor gauge group $SU(N)_{L+R+c}$ and 
the covariant derivative as
\begin{eqnarray}
 \D_\mu H_{ab}=\partial_{\mu}H_{ab}+i[A_\mu, H_{ab}], 
\quad H_{ab}\equiv H_a^\dagger H_b, \quad a,b=L,R.
\end{eqnarray}
In Appendix \ref{app2}, 
we will describe fully the procedure to derive the effective 
Lagrangian by substituting (\ref{eq:solU1_SY3}) and 
(\ref{eq:solU2_SY3}) and rewriting in terms of moduli 
fields $\hat{x}$ and $U$. 
Here we merely state the result:
\begin{align}
 \oper{L}_{1,{\rm eff}} & = \frac{v^2}{2m}
\Tr\biggl[\oper{D}_{\mu}\hat x\,
 \frac{\cosh(\oper{L}_{\hat x})-1}{\oper{L}_{\hat x}^2
\sinh(\oper{L}_{\hat x})}
\ln\biggl(\frac{1+\tanh(\oper{L}_{\hat x})}
{1-\tanh(\oper{L}_{\hat x})}\biggr)(\oper{D}^{\mu}\hat x) 
\nonumber \\
& +U^{\dagger}\oper{D}_{\mu}U\,
\frac{\cosh(\oper{L}_{\hat x})-1}{\oper{L}_{\hat x}
\sinh(\oper{L}_{\hat x})}
\ln\biggl(\frac{1+\tanh(\oper{L}_{\hat x})}
{1-\tanh(\oper{L}_{\hat x})}\biggr)(\oper{D}^{\mu}\hat x) 
\nonumber \\
&+\frac{1}{2}\oper{D}_{\mu}U^{\dagger}U\frac{1}
{\tanh(\oper{L}_{\hat x})}\ln\biggl(
\frac{1+\tanh(\oper{L}_{\hat x})}{1-\tanh(\oper{L}_{\hat x})}
\biggr)
(U^{\dagger}\oper{D}^{\mu}U)\biggr],
\label{eq:result}  
\end{align}
where 
\begin{equation}
{\cal L}_A(B)=[A,B]
\label{eq:lie_derivative}
\end{equation} 
is a Lie derivative with respect to $A$.
The covariant derivative $\oper{D}_\mu $  is defined by 
\beq
\oper{D}_\mu U = \p_\mu U + i \left[A_{\mu}, U\right].
\eeq

The above result suggests that the chiral fields $U(x^\mu)$ 
and hermitian fields $\hat x(x^{\mu})$ are 
in the adjoint representation of $SU(N)_{L+R}$. 
Let us now examine the transformation property of 
$U$ and $\hat x$ under the $SU(N)_{L+R}$ flavor gauge 
transformation on the domain wall background in order to 
demonstrate that they are in the adjoint representation. 
The domain wall solution only preserves the diagonal subgroup 
$SU(N)_{L+R+c}$. Eqs.(\ref{eq:transf_H}) and 
(\ref{eq:transf_sigma}) shows 
the fields transform under the $SU(N)_{L+R+c}$ 
transformations ${\cal U}$ as 
\begin{equation}
H'_L={\cal U} H_L {\cal U}^\dagger, \quad 
H'_R={\cal U} H_R {\cal U}^\dagger, \quad 
\Sigma'={\cal U} \Sigma {\cal U}^\dagger. 
\end{equation} 
Eqs.(\ref{eq:solU1_SY}) and (\ref{eq:solU2_SY}) show that 
\begin{equation}
S'={\cal U} S {\cal U}^\dagger, 
\quad e^{\phi'}={\cal U}e^\phi {\cal U}^\dagger, \quad 
\Omega'={\cal U}\Omega {\cal U}^\dagger.
\end{equation}
The complex moduli $e^\phi$ is decomposed into hermitian part 
$e^{\hat x}$ and unitary part $U$ in Eq.(\ref{eq:xu_decomp}). 
Since we can express 
$e^{2\hat x}=e^{\phi} e^{{\phi}^\dagger}$, 
and $U=e^{-\phi}e^{\hat x}$, 
we find that they transform as adjoint representations 
\begin{equation}
e^{2\hat x'}
={\cal U}e^{2\hat x} {\cal U}^\dagger, 
\quad U'={\cal U} U {\cal U}^\dagger. 
\end{equation}

By expanding (\ref{eq:result}), we here illustrate nonlinear 
interactions of $\hat{x}$ up to fourth orders in the 
fluctuations $\hat x$ and $U-{\bf 1}$ 
\begin{eqnarray}
\Lag_{1,\rm eff}&=&\frac{v^2}{2m}
\Tr\Bigr(\oper{D}_{\mu}U^{\dagger}\oper{D}^{\mu}U
+\oper{D}_{\mu}\hat x\oper{D}^{\mu}\hat x 
+U^{\dagger}\oper{D}_{\mu}U\comm{\hat x}{\oper{D}^{\mu}\hat x}
\nonumber \\
&&
-{1 \over 12}\comm{\oper{D}_{\mu}\hat x}{\hat x}
\comm{\hat x}{\oper{D}^{\mu}\hat x}
+\frac{1}{3}[\oper{D}_{\mu}U^\dagger U, \hat x]
[\hat x, U^\dagger \oper{D}^{\mu}U]
+\cdots \Bigr).
\label{eq:4thorder_nonlinear}
\end{eqnarray}

\begin{figure}[ht]
\begin{center}
\includegraphics[width=8cm]{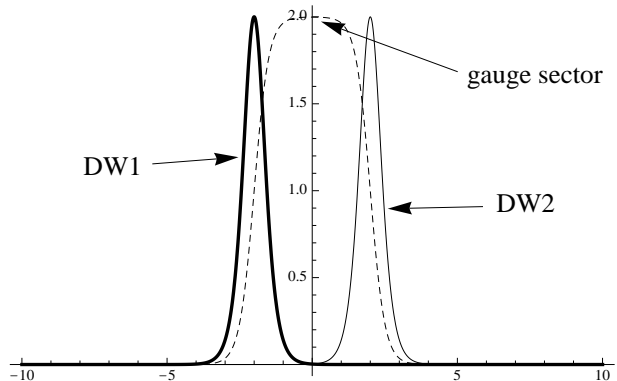}
\caption{The wave functions of the zero modes. 
DW1 and DW2 stand for the wave functions 
of the massless matter fields of the $i=1$ 
domain wall and $i=2$ domain wall, respectively 
for strong gauge coupling limit $g_i=\infty$ and 
$m_i=1$. 
The gauge fields are localized between the domain walls.}
\label{fig:schematic}
\end{center}
\end{figure}

Similarly to Eq.(\ref{eq:integral4Dcoupling}), we 
can define the (3+1)-dimensional non-Abelian gauge 
coupling $e_{4}$ 
by integrating (\ref{Lag_gChi2}) and find 
\beq
\frac{1}{2e_{4}^2} 
=\int dy \frac{1}{2e^2(\Sigma)} 
= \lambda(y_2-y_1)\,,
\label{eq:4d_NA_gauge_coupling}
\eeq
where $y_i$ is the wall position for the $i$-th domain wall 
sector. 
Summarizing, we obtain the following effective Lagrangian
\beq
\Lag_{\rm eff} = 
\Lag_{1, {\rm eff}} + 
\Lag_{2, {\rm eff}}- \frac{1}{2e_{4}^2}
\Tr \Bigl[G_{\mu\nu}G^{\mu\nu}\Bigr],
\label{eq:full_eff_Lag}
\eeq
where $\Lag_{2, {\rm eff}}$ is given in (\ref{eq:effLagAbelian}).
This is the main result of this paper.
We have succeeded in constructing the low energy 
effective theory in which the matter fields 
(the chiral fields) and the non-Abelian gauge 
fields are localized with the non-trivial interaction.
We show the profile of "wave functions" of localized 
massless gauge field and massless matter fields 
as functions of the coordinate $y$ of the extra 
dimension in Fig.~\ref{fig:schematic}.

As is seen from Eq.(\ref{eq:4thorder_nonlinear}), 
the flavor gauge symmetry $SU(N)_{L+R+c}$ is further 
(partly) broken and the corresponding gauge field 
$A_\mu$ becomes massive, when the fluctuation 
$\phi = e^{\hat x} U$ develops non-zero vacuum expectation values.
Especially, $\hat x$ is interesting because 
its non-vanishing (diagonal) values of the fluctuation 
has the physical meaning as the separation 
between walls away from the coincident case. 
For instance, if all the walls are separated, 
$SU(N)_{L+R+c}$ is spontaneously broken to
the maximal $U(1)$ subgroup $U(1)^{N-1}$. 
However, if $r$ walls are still coincident 
and all other walls are separated, we have an unbroken 
gauge symmetry $SU(r)\times U(1)^{N-r+1}$.
Then, a part of the pseudo-NG modes $\hat x$ turn to NG modes associated with
the further symmetry breaking $SU(N)_{L+R+c} \to SU(r) \times U(1)^{N-r+1}$, so that
the total number of zero modes is preserved\cite{EtFuNiOhSa}\footnote{In Ref.\citen{ShYu2} the authors argued that the non-Abelian clouds
spreading between walls become massive contrary to the results of Ref.\citen{EtFuNiOhSa}{}.}.
These new NG modes, called the non-Abelian cloud, spread between the separated domain walls\cite{EtFuNiOhSa}.
The flavor gauge fields eat the non-Abelian cloud and get masses which are proportional
to the separation of the domain walls. This is the Higgs mechanism in our model.
This geometrical understanding of the Higgs mechanism is quite similar to D-brane systems
in superstring theory. So our domain wall system provides a genuine prototype
of field theoretical D3-branes.

\bigskip
\section{Embedding into supersymmetric theory}\label{sec:SUSYE}
A crucial point to localize gauge field around domain wall 
is the coupling between scalar and gauge kinetic term. 
Such a coupling is naturally realized in (4+1)-dimensional
supersymmetric gauge theory \cite{OhSa}. 
This theory generally consists of hypermultiplet
part and vector multiplet part. 
The latter is specified by the so-called prepotential. 
In $(4+1)$-dimensional theory the prepotential generally 
allows up to cubic terms in vector multiplets\cite{Se}, 
which serves interactions among vector mutiplets such as 
\refer{Lag_gChi2}.

\smallskip
\subsection{Supersymmetric model}\label{SUSYE}
In embedding the model into supersymmetric gauge theories 
in $(4+1)$ dimensions, we will give non-Abelian global 
flavor symmetry $SU(N_i)_V$ for each copy ($i=1,2$) of 
the domain wall sector, instead of only one copy as in 
\refer{Lag_gChi} of the previous section. 
This contains the model \refer{Lag_gChi} as a limiting 
case of $N_2\to1$, and may offer more general situation 
phenomenologically. 
To formulate supersymmetric gauge theories, we need to 
introduce $Y_i$ as auxiliary fields of $U(N_i)_c$ 
vector multiplet, and $\Phi_i$ and ${\cal Y}_i$ as 
adjoint scalar fields and auxiliary fields of $SU(N_i)_V$ 
vector multiplet. 
As bosonic fields of theories with eight supercharges, 
we also need to double the scalar fields $H_i$, by 
 introducing another set 
$\tilde{H}^\dagger_i=(\tilde{H}_{iL}^\dagger,
\tilde{H}_{iR}^\dagger)$ with masses 
$(m_i{\bf 1}_{N_i}, -m_i{\bf 1}_{N_i})$. 
They are in the same representations as $H_i$ under 
$U(N_i)_c$ and $U(1)_{iA}$. 
Explicit charge assignments for hypermultiplets matter 
fields and adjoint scalar fields 
are summarized in Table \ref{table1}. 
\begin{table}[th]
\begin{center}
\begin{tabular}{c|cccc}
\hline
        & $U(N_i)_c$ & $U(1)_{iA}$ & $SU(N_i)_V$ & \mbox{mass} \\ \hline
$H_{iL}$ & $\Box$     & 1        & $\Box$         & $m_i{\bf 1}_{N_i}$ \\
$H_{iR}$ & $\Box$     & 1        & $\Box$         & $-m_i{\bf 1}_{N_i}$ \\
$\tilde{H}_{iL}$ & $\bar{\Box}$ & $-1$ & $\bar{\Box}$ & $m_i{\bf 1}_{N_i}$ \\
$\tilde{H}_{iL}$ & $\bar{\Box}$ & $-1$ & $\bar{\Box}$ & $-m_i{\bf 1}_{N_i}$ \\
$\Sigma_i$ & {adj} & 0 & ${\bf 1}$ & 0 \\
$\Phi_i$ & ${\bf 1}$ & 0 & adj & 0 \\
\hline
\end{tabular}
\end{center}
\caption{Quantum numbers of hypermultiplets $(H_i,\tilde{H}_i)$, $\Sigma_i$ and $\Phi_i$.}
\label{table1}
\end{table}
The resultant supersymmetric Lagrangian is written as 
\begin{eqnarray}
{\cal L}&=&
a_{\alpha\beta}(\Sigma)\left(-{1\over 4} F_{MN}^\alpha 
F^{\beta MN}
+{1\over 2} \mathcal{D}_M \Sigma^\alpha 
\mathcal{D}^M \Sigma^\beta 
+{1\over 2}Y^{\alpha}Y^{\beta }
\right) - c_{\alpha } Y^{\alpha } \nonumber \\
&&+
 \sum_{i=1}^2{\rm Tr}{\Big \{}\left(\tilde{\cal D}_M H_{iL} 
\tilde{\cal D}^M H_{iL}^\dagger
 +\tilde{\cal D}_M \tilde{H}_{iL} \tilde{\cal D}^M 
\tilde{H}_{iL}^\dagger 
 +(L \leftrightarrow R)\right) \nonumber \\
&&\quad \qquad \qquad -V_{iF}+\mathcal{L}_{iY}
+{\cal L}_{i\rm CS}+{\cal L}_{i\rm fermion}{\Big \}}\,,
\label{eq:5DLagrangian}
\end{eqnarray}
where
\begin{eqnarray}
\mathcal{L}_{iY}&=&{\rm Tr}\Bigl[H_{iL}^\dagger Y_i H_{iL}
-H_{iL}^\dagger H_{iL} {\cal Y}_i
-\tilde{H}_{iL}Y_i \tilde{H}_{iL}^\dagger 
+{\cal Y}_i\tilde{H}_{iL}\tilde{H}_{iL}^\dagger
+(L \leftrightarrow R)
\Bigr], \\
V_{iF}&=&{\rm Tr}{\Big [}|\Sigma_i H_{iL}-H_{iL}
(\Phi_i+m_i{\bf 1}_{N_i})|^2 
+ |\tilde{H}_{iL} \Sigma_i-(\Phi_i+m_i{\bf 1}_{N_i})
\tilde{H}_{iL} |^2 
\nonumber \\
&&+|\Sigma_i H_{iR}-H_{iR}(\Phi_i-m_i{\bf 1}_{N_i})|^2 
+ |\tilde{H}_{iR} \Sigma_i-(\Phi_i-m_i{\bf 1}_{N_i})
\tilde{H}_{iR} |^2
{\Big ]},
\label{F-pot}
\end{eqnarray}
where $\alpha, \beta\cdots$ denote all gauge groups and 
their generators collectively. 
We label them with the ordering 
\begin{eqnarray}\label{eq:colflav}
\alpha, \beta = 0_1,I_1,A_1;0_2,I_2,A_2\,,
\end{eqnarray}
where $0_i$ denotes $U(1)_i$ parts of $U(N_i)_c$ gauge 
group, while $I_i=1, \cdots, N_i^2-1$ are color indices
 of $SU(N_i)_c$ and 
$A_i=1, \cdots, N_i^2-1$ denotes flavor indices of 
$SU(N_i)_V$ gauge group. 
The scalar fields $\Sigma^\alpha$ and auxiliary 
fields $Y^\alpha$ are explicitly written by
\begin{eqnarray}
 \Sigma^\alpha 
&=&(\Sigma^{0_1},\Sigma^{I_1},\Phi^{A_1};\Sigma^{0_2},\Sigma^{I_2},\Phi^{A_2})\,,\\  
 Y^\alpha 
 &=&(Y^{0_1}, Y^{I_1},\mathcal{Y}^{A_1};Y^{0_2}, Y^{I_2},\oper{Y}^{A_2})\,,
\end{eqnarray}
and similarly the field strength $F_{MN}^\alpha$ and gauge 
field $W_{M}^\alpha$ are written by
\begin{align}
F_{MN}^\alpha &=(F_{MN}^{0_1}, F_{MN}^{I_1},G_{MN}^{A_1}; F_{MN}^{0_2}, F_{MN}^{I_2},  G_{MN}^{A_2})\,,\\ 
W_{M}^\alpha &=(W_{M}^{0_1}, W_{M}^{I_1},A_{M}^{A_1};W_{M}^{0_2}, W_{M}^{I_2}, A_{M}^{A_2})\,.
\end{align}
We adopt the convention of $U(N_i)_c$ and $SU(N_i)_V$ matrices such as
\begin{eqnarray}
 \Sigma_i&=&\Sigma^{0_i}\mathbf{1}_{N_i}+\Sigma^{I_i}T^{I_i}\,,\quad {\rm Tr}(T^{I_i} T^{J_i}) ={1 \over 2}\delta^{I_iJ_i},\\
 \Phi_i&=&\Phi^{A_i}T^{A_i}\,,\quad {\rm Tr}(T^{A_i} T^{B_i}) ={1 \over 2}\delta^{A_iB_i},\quad (\mbox{no~sum~for}~i ).
\end{eqnarray}
Covariant derivatives for $H_{iL}$ and $H_{iR}$ are given as 
\refer{eq:covd1} and \refer{eq:covd2} with identical 
definition for $\tilde{H}_{iL}^\dagger$ $\tilde{H}_{iR}^\dagger$. 
Covariant derivatives of $\Sigma^{I_i}, \Phi^{A_i}$ 
are defined as the adjoint representation.
We will not display the Chern-Simons term ${\cal L}_{i\rm CS}$ 
and the fermionic term ${\cal L}_{i\rm fermion}$, since we do 
not need them for our analysis. 

Functions $a_{\alpha\beta}(\Sigma)$ are gauge coupling 
functions, which are given as second derivative of 
the prepotential 
\begin{eqnarray}
a(\Sigma)=\sum_{i=1}^2\biggl[\frac{1}{2\hat g_i^2}
(\Sigma^{0_i})^2+{1 \over 2g^2_i}(\Sigma^{I_i})^2
+\frac{\lambda_i}{2}\left({\Sigma^{0_1} \over m_1}
-{\Sigma^{0_2} \over m_2}\right)(\Phi^{A_i})^2\biggr]\,, 
\label{eq:prepotential}
\end{eqnarray}
\begin{eqnarray}
a_{\alpha\beta}(\Sigma)
={\partial^2 a(\Sigma) \over \partial \Sigma^\alpha 
\partial \Sigma^\beta}\,.
\label{eq:2nd_der_prepotential}
\end{eqnarray}
From the above prepotential, we see the coupling constants 
of $U(1)_{i}$ and $SU(N_i)_c$ are given by 
$\hat g_i$ and $g_i$, respectively\footnote{
The $U(1)_{i}$ coupling is in principle unrelated to the 
$SU(N_i)_c$ coupling. In section~\ref{sec:chiral model}, 
we made a simplifying 
assumption $\hat g_i=g_i/\sqrt{2N_i}$, which allows 
simple solutions. 
}. We denote the coupling function of $SU(N_i)_V$ 
corresponding to $\Sigma^\alpha=\Phi^{A_i}$ 
and $\Sigma^\beta=\Phi^{B_i}$ 
as $e_i(\Sigma)$, 
%
%
\begin{eqnarray}
{1 \over e^2_i(\Sigma)} 
= \lambda_i\left({\Sigma^{0_1} \over m_1}
-{\Sigma^{0_2} \over m_2}\right)
\,,
\end{eqnarray}
but will suppress the argument $\Sigma$ to write $e_i$ 
in the following. 

The constants $c_{\alpha}$ are coefficients of the 
Fayet-Iliopoulos (FI) terms, allowed 
to be non-zero only for the $U(1)$ part of the gauge 
groups\footnote{
The FI parameters $c_{0_i}$ are related to the parameters 
$v^2_i$ in Eqs.(\ref{eq:Lag_chi_sum})-(\ref{eq:YM_Higgs_Lag}) 
in section~\ref{sec:chiral model} as $c_{0_i}=N_i v^2_i$. 
}
\begin{equation}
c_{\alpha}Y^{\alpha}=c_{0_1}Y^{0_1}+c_{0_2}Y^{0_2}\,.
\end{equation}
We have assumed both the FI parameters $c_{0_1}$ and $c_{0_2}$ 
to be positive in the same direction in $SU(2)_R$, which is 
chosen to be along the third component. 
In this setup, the $\tilde H$ fields will vanish in the 
classical solution. Moreover, they do not contribute to 
the desired order of effective Lagrangian. 
Similarly we have neglected the auxiliary fields $Y$ 
other than the third component in $SU(2)_R$, 
and we have denoted as $Y^{\alpha}$. 
Hence we can call the potential after eliminating the 
auxiliary fields $Y$'s to be D-term potential. 

The F-term potential $V_{iF}$ can be worked out from 
the following superpotential 
\begin{eqnarray}
 W_i&=&{\rm Tr}\left[\left\{\Sigma_i H_{iL}-H_{iL}
(\Phi+m_i) \right\}\tilde{H}_{iL}
+\left\{\Sigma_iH_{iR}-H_{iR}(\Phi-{m_i}) \right\}
\tilde{H}_{iR}\right], 
\label{eq:superpotential}
\end{eqnarray}
where we restored the tilde fields $\tilde H$'s to 
facilitate writing the superpotential. 
After eliminating the auxiliary fields $F$'s, and 
with the use of 
\begin{eqnarray}
 V_{iF}=-{\cal L}_{iF}=|F_{iL}|^2+|F_{iR}|^2
+|\tilde{F}_{iL}|^2+|\tilde{F}_{iR}|^2\,,
\end{eqnarray}
we have (\ref{F-pot}).

Finally, let us work out explicit forms of the D-term potential 
$V_D$. Collecting terms containing the auxiliary fields $Y$'s, 
we obtain 
\begin{eqnarray}\label{eq:vdd}
-V_{D}&=&\sum_{i=1}^2{\Bigg [}{1\over 2\hat g_i^2}
(Y^{0_i})^2+{1 \over 2g_i^2}(Y^{I_i})^2
+r^{I_i}Y^{I_i}+(r^{0_i}-c_{0_i})Y^{0_i}
+{1 \over 2e_i^2}({\cal Y}^{A_i})^2
+s^{A_i} {\cal Y}^{A_i} \nonumber \\
&&+\lambda_i\left({ Y^{0_1} \over m_1}
-{Y^{0_2} \over m_2}\right)\Phi^{A_i}
 {\cal Y}^{A_i}{\Bigg ]}, \label{Dp}
\end{eqnarray}
where
\begin{eqnarray}
r_i&=&H_{iL}H_{iL}^\dagger-\tilde{H}_{iL}^\dagger 
\tilde{H}_{iL}+H_{iR}H_{iR}^\dagger
-\tilde{H}_{iR}^\dagger\tilde{H}_{iR}\,, \\
s_i&=&-H_{iL}^\dagger H_{iL}+\tilde{H}_{iL}
\tilde{H}_{iL}^\dagger-H_{iR}^\dagger H_{iR}
-\tilde{H}_{iR}\tilde{H}_{iR}^\dagger\,,
\end{eqnarray}
are Hermitian matrices, with the decomposition
\begin{eqnarray}
r_i=\frac{1}{N_i}r^{0_i}\mathbf{1}_{N_i}+2 r^{I_i} T^{I_i},\quad s_i=-{1 \over N_i}r^{0_i}\mathbf{1}_{N_i}+2s^{A_i} T^{A_i}\,.
\end{eqnarray}
We observe, that in the potential \refer{eq:vdd}, 
$Y^{I_i}$ do not couple to the rest of
auxiliary fields and can be easily eliminated. 
Having this done, we collect the $U(1)_i$ and 
$SU(N_i)_V$ terms into a matrix form labeled by 
 $\alpha,\beta = 0_1, 0_2, A_1, A_2$ 
\begin{equation}\label{eq:vd}
-V_D = -\frac{1}{2}\sum_{i=1}^2 g_i^2 (r^{I_i})^2
+\frac{1}{2}G_{\alpha\beta}Y^{\alpha}Y^{\beta}
+(r-c)_{\alpha}Y^{\alpha}\,,
\end{equation}
\begin{eqnarray}
(r-c)_{0_i}\equiv r^{0_i}-c_{0_i}\,,\quad (r-c)_{A_i}\equiv s^{A_i}\,.
\end{eqnarray}
Eliminating remaining auxiliary fields we obtain:
\begin{equation}
V_D = \frac{1}{2}\sum_{i=1}^2g_i^2 (r^{I_i})^2
+ \frac{1}{2}(G^{-1})^{\alpha\beta}(r-c)_{\alpha}(r-c)_{\beta}\,.
\end{equation}
Matrix $G=(G_{\alpha\beta})$ is explicitly given by
\begin{equation}\label{eq:g}
G = \begin{pmatrix}
\frac{1}{\hat g_1^2} & 0 & \frac{\lambda_1}{m_1}\Phi^{A_1} 
& \frac{\lambda_2}{m_1}\Phi^{A_2} \\
0 & \frac{1}{\hat g_2^2} & -\frac{\lambda_1}{m_2}\Phi^{A_1} 
& -\frac{\lambda_2}{m_2}\Phi^{A_2} \\
\frac{\lambda_1}{m_1}\Phi^{B_1} 
& -\frac{\lambda_1}{m_2}\Phi^{B_1} 
& \frac{1}{e_1^2}\delta^{A_1B_1} & 0 \\
\frac{\lambda_2}{m_1}\Phi^{B_2} 
& -\frac{\lambda_2}{m_2}\Phi^{B_2} & 0 
& \frac{1}{e_2^2}\delta^{A_2B_2}
\end{pmatrix}\,,
\end{equation} 
with the inverse
\begin{gather}\label{eq:ginverse}
G^{-1} = \frac{1}{1-\tilde g^2\tilde\Phi^2}\times \nonumber\\
\begin{pmatrix}
\hat g_1^2-\hat g_1^2\hat g_2^2m_1^2\tilde\Phi^2 
& -\hat g_1^2\hat g_2^2m_1m_2\tilde\Phi^2 
& -\hat g_1^2e_1m_2\tilde\Phi^{A_1} 
& -\hat g_1^2e_2m_2\tilde\Phi^{A_2} \\
-\hat g_1^2\hat g_2^2m_1m_2\tilde\Phi^2 
& \hat g_2^2-\hat g_1^2\hat g_2^2m_2^2\tilde\Phi^2 
& \hat g_2^2e_1m_1\tilde\Phi^{A_1} 
& \hat g_2^2e_2m_1\tilde\Phi^{A_2} \\
-\hat g_1^2e_1m_2\tilde\Phi^{B_1} 
& \hat g_2^2e_1m_1\tilde\Phi^{B_1} 
& e_1^2\delta^{A_1B_1}-e_1^2\tilde g^2\tilde\Phi^{A_1B_1} 
& \tilde g^2e_1e_2\tilde\Phi^{A_2}\tilde\Phi^{B_1} \\
-\hat g_1^2e_2m_2\tilde\Phi^{B_2} &  \hat g_2^2e_2m_1\tilde\Phi^{B_2} 
& \tilde g^2e_1e_2\tilde\Phi^{A_1}\tilde\Phi^{B_2} 
& e_2^2\delta^{A_2B_2}-e_2^2\tilde g^2\tilde\Phi^{A_2B_2}
\end{pmatrix}\,,
\end{gather}
where we abbreviated:
\begin{align}
\label{eq:tl1} \tilde g^2 & 
= \hat g_1^2m_2^2+\hat g_2^2m_1^2\,, \\
\label{eq:tl2} \tilde \Phi^{A_i} & 
= \frac{\lambda_ie_i}{m_1m_2}\Phi^{A_i}\,, \\
\label{eq:tl3} \tilde \Phi^2 & 
= \tilde \Phi^{A_1}\tilde\Phi^{A_1}
+\tilde \Phi^{A_2}\tilde\Phi^{A_2}\,, \\
\label{eq:tl4} \tilde \Phi^{A_iB_i} & 
= \tilde\Phi^2\delta^{A_iB_i}
-\tilde\Phi^{A_i}\tilde\Phi^{B_i} \,.
\end{align}

\smallskip
\subsection{Positivity of Potential}
The F-term potential (\ref{F-pot}) is manifestly positive. 
The D-term potential \refer{eq:vd} is positive definite 
under certain conditions. 
To find the condition we shall decompose \refer{eq:vd} to:
\begin{align}
 V_D &= V_{1D}+V_{2D},\\
 V_{1D}&=  \frac{1}{2}g_1^2 (r^{I_1})^2
+\frac{1}{2}g_2^2 (r^{I_2})^2, \\
V_{2D}&= \frac{1}{2}(G^{-1})^{\alpha\beta}
(r-c)_{\alpha}(r-c)_{\beta}\,.
\end{align}
It is clear that the $V_{1D}$ is positive definite by itself. 
Therefore we can only focus on $V_{2D}$, which is positive 
if and only if $G$ is positive definite.
  
It is easy to recognize that positivity of $G$ is manifest 
once the adjoint scalars vanish $\Phi_i=0$.
Nevertheless, it is instructive and assuring if we consider 
the potential as well as the BPS equations keeping the 
adjoint scalars $\Phi_i$ nonzero.

To ascertain positivity of $G$ we need to compute its 
eigenvalues. This is most easily done by looking at its 
determinant (We leave the derivation of this result to 
the Appendix \ref{app:det}):
\begin{equation}
\label{eq:detg}\det G = \Bigl[\frac{1}{\hat g_1^2\hat g_2^2}
-\left(\frac{m_1^2}{\hat g_1^2}+\frac{m_2^2}{\hat g_2^2}\right)
\tilde\Phi^{2}\Bigr]
\Bigl(\frac{1}{e_1^2}\Bigr)^{N_1}\Bigl
(\frac{1}{e_2^2}\Bigr)^{N_2}\,.
\end{equation}
Requiring $\det G>0$, we have
\begin{eqnarray}
\label{detG} \frac{1}{\hat g_1^2\hat g_2^2}
-\Bigl(\frac{m_1^2}{\hat g_1^2}
+\frac{m_2^2}{\hat g_2^2}\Bigr)\tilde\Phi^2  > 0. 
\end{eqnarray}
In Appendix \ref{app:det} we show that this condition is 
both necessary and sufficient to ensure positivity of 
matrix $G$ in Eq.(\ref{eq:g}).

\smallskip
\subsection{BPS equations}\label{sc:3}

Let us denote the codimension of the domain wall as $y$. 
Since we assume Lorentz invariance for other dimensions, 
we obtain gauge field to vanish for component other than $y$. 

The energy density ${\cal H}$ for domain walls is given by 
\begin{eqnarray}
{\cal H}&=&
\frac{1}{2}G_{\alpha\beta}\oper{D}_y\Sigma^\alpha
\oper{D}_y\Sigma^\beta 
+\frac{1}{2}(G^{-1})^{\alpha\beta}(r-c)_{\alpha}
(r-c)_{\beta} \\
&+&\sum_{i=1}^2{\rm Tr}
\left\{
\bigl(\tilde{\oper{D}}_y H_{iL}
\tilde{\oper{D}}_yH_{iL}^\dagger
+\tilde{\oper{D}}_yH_{iR}\tilde{\oper{D}}_yH_{iR}^\dagger 
+\tilde{\oper{D}}_y\tilde H_{iL}^{\dagger}
\tilde{\oper{D}}_y\tilde H_{iL}
+\tilde{\oper{D}}_y\tilde H_{iR}^{\dagger}
\tilde{\oper{D}}_y\tilde H_{iR})+V_{iF}\right\}\,,
\nonumber
\label{eq:energy_density}
\end{eqnarray}
where color-flavor indices $\alpha, \beta$ span all values 
as in Eq.\refer{eq:colflav} and we have incorporated 
color sector $\alpha = I_1, I_2$ into 
the definition of matrix $G$ for brevity. 
Accordingly, we have incorporated the definition, 
$(r-c)_{I_i}=r^{I_i}$. 
Since there is no mixing of color sector with the rest, 
the inverse is calculated 
trivially and non-color part remains the same as in 
\refer{eq:ginverse}. 

Now we observe that the mixing due to the cubic prepotential 
occurs only in the kinetic term and potential of the 
vector multiplets. 
Moreover, they appear as $G$ and $G^{-1}$ respectively. 
Therefore the cross term coming out of the Bogomol'nyi 
completion has no dependence on the metric $G$. 
This fact implies that the cancellation of cross terms 
to give 
topological charge goes through unaffected by the mixing of 
the vector multiplets. 

More explicitly, we obtain the Bogomol'nyi completion as 
\begin{eqnarray}
{\cal H}
&=&
\frac{1}{2}
\biggl(G_{\alpha\gamma}\oper{D}_y\Sigma^\gamma
+(r-c)_\alpha\biggr)(G^{-1})^{\alpha\beta}
\biggl(G_{\beta\delta}\oper{D}_y\Sigma^\delta
+(r-c)_\beta\biggr)
\nonumber \\
&+&
\sum_{i=1}^2{\rm Tr}\Bigl[
\bigl|\tilde{\oper{D}}_yH_{iL}+\Sigma_i H_{iL} 
-H_{iL}\left(\Phi_i+m_i\mathbf{1}_{N_i}\right)\bigr|^2 
\nonumber \\
&&+\bigl|\tilde{\oper{D}}_y\tilde H_{iL}^\dagger
-\Sigma_i \tilde H_{iL}^\dagger 
+\tilde H_{iL}^\dagger\left(\Phi_i+m_i\mathbf{1}_{N_i}
\right)\bigr|^2
\nonumber \\
&&
+\bigl|\tilde{\oper{D}}_yH_{iR}
+\Sigma_i H_{iR} -H_{iR}\left(\Phi_i-m_i\mathbf{1}_{N_i}
\right)\bigr|^2 \nonumber \\
&&+\bigl|\tilde{\oper{D}}_y\tilde H_{iR}^\dagger
-\Sigma_i \tilde H_{iR}^\dagger 
+\tilde H_{iR}^\dagger\left(\Phi_i
-m\mathbf{1}_{N_i}\right)\bigr|^2\Bigr]
\nonumber \\
&
-
&
\sum_{i=1}^2\partial_y\mathrm{Tr}\Bigl[\Sigma_{i}r_{i}
+\Phi_i s_i-m_i(H_{iL}^{\dagger}H_{iL}-H_{iR}^{\dagger}H_{iR}
-\tilde H_{iL}\tilde H_{iL}^{\dagger}+\tilde H_{iR}
\tilde H_{iR}^{\dagger})\Bigr] \nonumber \\
&
+&
c_\alpha \partial_y {\rm Tr}\Sigma^{\alpha}\,. 
\label{eq:bogomolnyi}
\end{eqnarray}
The last term gives the usual Bogomol'nyi bound and becomes 
the topological charge. 
The line before that is the total derivative 
which give vanishing contribution 
for an infinite line $-\infty < y < \infty$. 

BPS equations for $H$'s and $\tilde H$'s of 
hypermultiplets are 
\begin{eqnarray}
&&\tilde{\oper{D}}_yH_{iL}+\Sigma_i H_{iL} 
-H_{iL}\left(\Phi_i+m_i\mathbf{1}_{N_i}\right)=0\,, 
\label{eq:bpsh1}\\
&&\tilde{\oper{D}}_y\tilde H_{iL}^\dagger
-\Sigma_i \tilde H_{iL}^\dagger 
+\tilde H_{iL}^\dagger\left(\Phi_i
+m_i\mathbf{1}_{N_i}\right)=0\,, \label{eq:bpsh2}\\
&&\tilde{\oper{D}}_yH_{iR}
+\Sigma_i H_{iR} -H_{iR}\left(\Phi_i
-m_i\mathbf{1}_{N_i}\right)=0\,,  \label{eq:bpsh3}\\
&&\tilde{\oper{D}}_y\tilde H_{iR}^\dagger
-\Sigma_i \tilde H_{iR}^\dagger
+\tilde H_{iR}^\dagger\left(\Phi_i
-m_i\mathbf{1}_{N_i}\right)=0\,. \label{eq:bpsh4}
\end{eqnarray}
BPS equations for vector multiplets are 
\begin{eqnarray}
G_{\alpha\beta}\oper{D}_{y}\Sigma^\beta+(r-c)_\alpha=0\,.
\end{eqnarray}
More explicitly, 
\begin{eqnarray}
&&\frac{1}{\hat g_i^2}\partial_{y}\Sigma^{0_i}
+\sum_{j,k=1}^2\frac{\lambda_j\varepsilon_{ik}m_k}
{m_1m_2}\Phi^{A_j}\oper{D}_{y}\Phi^{A_j}
+r^{0_i}-c_{0_i} = 0\,, \label{eq:bpsv1} \\
&&\frac{1}{g_i^2}\oper{D}_{y}\Sigma^{I_i}+r^{I_i} = 0\,, 
\label{eq:bpsv2}\\
&&
{1 \over e_i^2}{\cal D}_y\Phi^{A_i}
+\sum_{j,k=1}^2\frac{\lambda_i\varepsilon_{jk}m_k}
{m_1m_2}\Phi^{A_i}\partial_{y}\Sigma^{0_j}
+s^{A_i} = 0\,. \label{eq:bpsv3}
\end{eqnarray}

We can easily solve the BPS equation for hypermultiplets, 
by using the moduli matrix approach. 
We define $S_{ic}, S_{iF}$ and $\psi_i$ as 
\begin{eqnarray}
&&\Sigma_i(y)+i W_{iy}(y)
=S_{ic}^{-1}(y)\partial_y S_{ic}(y)
+\frac{1}{2}\partial_y\psi_i(y)\,, 
\label{eq:master_variableSC} \\
&&\Phi_i(y)+\mathrm{i}A_{iy}(y)
=S_{iF}(y)\partial_y S_{iF}^{-1}(y)\,.
\label{eq:master_variableSF}
\end{eqnarray}
Then the hypermultiplets BPS equations 
(\ref{eq:bpsh1})-(\ref{eq:bpsh4}) are solved by the 
constant moduli matrices $H_{iL}^{0}$ and $H_{iR}^{0}$ 
\begin{eqnarray}
&&H_{iL} 
= e^{-\psi_i/2}S_{ic}^{-1} H_{iL}^{0} 
S_{iF}^{-1} e^{m_iy}\,, 
\label{eq:hyper_bps_sol1l}  \\
&&H_{iR} 
= e^{-\psi_i/2}S_{ic}^{-1} H_{iR}^{0} 
S_{iF}^{-1} e^{-m_iy}\,, 
\label{eq:hyper_bps_sol1r}
\end{eqnarray}
where $S_{ic}, S_{iF} \in SL(N_i,\mathbb{C})$.
The hypermultiplet fields $\tilde{H}_{iL}$ and 
$\tilde{H}_{iR}$ do not contribute
to domain wall solution and they are therefore vanishing.
We write down (\ref{eq:bpsv1})-(\ref{eq:bpsv3}) 
in terms of the gauge invariant fields 
\begin{eqnarray}
\Omega_{ic} = S_{ic} S_{ic}^\dagger\,, \quad
\Omega_{iF} = S_{iF}^\dagger S_{iF}\,, \quad 
\eta_{i} = \frac{1}{2}(\psi_i+\psi_i^*)\,.
\end{eqnarray} 
The adjoint scalar fields of the vector multiplets are 
given by 
\begin{eqnarray}
 \Sigma_i
&=& 
\frac{1}{2} S_{ic}^{-1}( \partial_y\Omega_{ic} 
\Omega_{ic}^{-1}) S_{ic} +\frac{1}{2}\partial_y\eta_i \,,
\label{eq:Sigma_by_omega}
\\
\Phi_i
&=& 
-\frac{1}{2} S_{iF}^{\dagger -1}( \partial_y\Omega_{iF} 
\Omega_{iF}^{-1}) S_{iF}^\dagger \,.
\label{eq:Phi_by_omega}
\end{eqnarray} 
Also, we have
\begin{eqnarray}
&&D_y  \Sigma_i = \partial_y \Sigma_i + i 
\left[ W_{iy}, \Sigma_i \right] =
 \frac{1}{2}S_{ic}^{-1}\partial_y( \partial_y \Omega_{ic} 
\Omega_{ic}^{-1}) S_{ic} +\frac{1}{2}\partial_y^2\eta_i \,,\\
&&D_y \Phi_i = \partial_y \Phi_i 
+ i \left[ A_{iy},\Phi_i\right] 
= - \frac{1}{2}S_{iF}^{\dagger -1}
\partial_y( \partial_y \Omega_{iF} \Omega_{iF}^{-1}) 
S_{iF}^\dagger\,.
\end{eqnarray}

BPS equations for vector multiplets 
(\ref{eq:bpsv1})-(\ref{eq:bpsv3}) can be now rewritten as 
the following master equations:
\begin{align}
& \frac{1}{2\hat g_i^2}\partial_y^2\eta_i 
+ \frac{\varepsilon_{ik}m_k}{2m_1m_2}\mathrm{Tr}
\Bigl[\lambda_j(\partial_y\Omega_{jF}
\Omega_{jF}^{-1})^2\Bigr] \nonumber\\
& \qquad = c_{0_i} -e^{-\eta_i}\mathrm{Tr}
\Bigl[\bigl(H_{iL}^0\Omega_{iF}^{-1}
H_{iL}^{0\dagger}e^{2m_iy}+
H_{iR}^0\Omega_{iF}^{-1}H_{iR}^{0\dagger}
e^{-2m_iy}\bigr)\Omega_{ic}^{-1}\Bigr]\,, 
\label{eq:masterEqe}
\\
& \frac{1}{g_i^2}\partial_y(\partial_y\Omega_{ic}
\Omega_{ic}^{-1}) \nonumber \\
& \qquad = -e^{-\eta_i}\Bigl<\bigl(H_{iL}^0
\Omega_{iF}^{-1}H_{iL}^{0\dagger}e^{2m_iy}+
H_{iR}^0\Omega_{iF}^{-1}H_{iR}^{0\dagger}
e^{-2m_iy}\bigr)\Omega_{ic}^{-1}\Bigr>\,, 
\label{eq:masterEqC}
\\
& \frac{\lambda_i}{m_1m_2}\partial_{y}
(\partial_y(\eta_1m_2-\eta_2m_1)
\partial_y\Omega_{iF}\Omega_{iF}^{-1})\nonumber \\
& \qquad = -e^{-\eta_i}\Bigl
<\bigl(H_{iL}^{0\dagger}\Omega_{ic}^{-1}H_{iL}^{0}e^{2m_iy}+
H_{iR}^{0\dagger}\Omega_{ic}^{-1}H_{iR}^{0}
e^{-2m_iy}\bigr)\Omega_{iF}^{-1}\Bigr>\,.
\label{eq:masterEqF}
\end{align} 
Here we have used a notation
\begin{eqnarray}
\left<X \right> \equiv X - \frac{{\rm Tr}[X]}{N} {\bf 1}_N.
\end{eqnarray} 
We make a comment about a possibility of additional moduli. 
At present we cannot say definitely if there are additional 
moduli other than the moduli matrices $H_{0}$'s, since 
we cannot solve these master equations. 
We have several clues at hand. 
The BPS equations for domain walls and other solitons 
in gauge theories with scalar fields in the fundamental 
representations are in the Higgs phase where all the 
gauge symmetries are broken in the vacuum. 
In that situation, we learned that all the moduli are 
contained in the moduli matrix. 
On the other hand, instantons are solitons in the pure 
Yang-Mills theory without scalar fields, where gauge 
symmetry is unbroken in the vacuum. In this case, moduli 
reside in the BPS equation for gauge fields. 
In our present case, unbroken gauge symmetry $SU(N_i)_{c+V}$ 
remains. 
This feature is indicative of additional moduli coming from 
the vector multiplet. 

Irrespective of the possible additional moduli, we 
can demonstrate that the BPS equations admit the 
coincident wall solution. 
Since the hypermultiplet parts are already solved as in 
(\ref{eq:hyper_bps_sol1l})-(\ref{eq:hyper_bps_sol1r}),
our main task is to solve the master equations 
(\ref{eq:masterEqe})-(\ref{eq:masterEqF}) associated 
to the vector multiplet. 
In order to solve them explicitly, we take strong gauge 
coupling limit $\hat g_i, g_i\rightarrow \infty$, 
where the master equations give just the algebraic 
constraints for $\Omega_{ic},\Omega_{iF}$ and $\eta_i$. 
In principle, they can be solved algebraically. 
Furthermore, Eq.(\ref{detG}) with the limit 
$g_i\rightarrow \infty$ tells us that positivity is 
maintained only if $\Phi_i$ vanishes. 
In the following we will, therefore, consider a special 
point in the solution space where
\begin{eqnarray}
\Phi_i=0, \quad i=1,2 \label{P0}
\end{eqnarray}
which implies from Eq.(\ref{eq:Phi_by_omega}) 
that $\Omega_{iF}$ are constant matrices. 
Then the differential equations 
(\ref{eq:masterEqC})-(\ref{eq:masterEqF}) reduce to the 
set of algebraic equations:
\begin{eqnarray}\label{eq:master_gsy1}
H_{iL}^0\Omega_{iF}^{-1}H_{iL}^{0\dagger}e^{2m_iy}+
H_{iR}^0\Omega_{iF}^{-1}H_{iR}^{0\dagger}e^{-2m_iy} 
= \frac{c_i}{N_i}e^{\eta_i}\Omega_{ic}\,,\\
\label{eq:master_gsy2}
H_{iL}^{0\dagger}\Omega_{ic}^{-1}H_{iL}^{0}e^{2m_iy}+
H_{iR}^{0\dagger}\Omega_{ic}^{-1}H_{iR}^{0}e^{-2m_iy} 
= \frac{c_i}{N_i}e^{\eta_i}\Omega_{iF}\,.
\end{eqnarray}
Notice, that for both sectors $i=1,2$ these equations 
are the same and do not couple to each other. 
We can, therefore, focus our discussion only 
on one sector, since all results are equivalent in both 
of them. So in the remaining discussion we will drop 
the index $i$ from all fields.

Now we consider moduli matrix for the coincident walls 
corresponding to the most symmetric point of the moduli 
space. 
\begin{eqnarray}
\left( H_{L}^{0}, H_{R}^{0} \right) 
= \left( {\bf 1}_N, {\bf 1}_N\right). 
\label{eq:fixedpoint_moduli}
\end{eqnarray}
Eqs.(\ref{eq:master_gsy1}) and (\ref{eq:master_gsy2}) 
show that these two constant matrices commute and 
only the product 
$\Omega_c\Omega_F = \Omega_F \Omega_c$ 
can be determined\footnote{
This is due to the special choice of the moduli matrix in 
Eq.(\ref{eq:fixedpoint_moduli}), since 
Eqs.(\ref{eq:hyper_bps_sol1l}) and (\ref{eq:hyper_bps_sol1r}) 
imply that only the product $S_{iF}S_{ic}$ 
can enter into the physical fields such as $H_{iL}, H_{iR}$. 
}
\begin{eqnarray}
e^{\eta} \Omega_c\Omega_F 
= \frac{N}{c} \left( e^{2my} + e^{-2my}\right){\bf 1}_N.
\end{eqnarray}
Since we have chosen the matrices 
$S_c, S_F$ in $SL(N,\mathbb{C})$, 
we find that det$(\Omega_c\Omega_F)=1$ and we can 
separate the $U(1)$ part. 
\begin{eqnarray}
e^{\eta} = \frac{N}{c} \left( e^{2my} + e^{-2my}\right), 
\quad 
 \Omega_c\Omega_F 
= {\bf 1}_N.
\end{eqnarray}
The $U(1)$ part gives the usual domain wall solution. 
Without affecting the physical quantities, we can 
choose $\Omega_{c}=1, S_{c}=1$, and finally we obtain 
the coincident wall solution for 
(\ref{eq:fixedpoint_moduli}) with (\ref{P0}) and with the wall position moduli $y_0$ (modifying (4.61) to $(H_L^0, H_R^0)
=(e^{-my_0}\mathbf{1}_N, e^{my_0}\mathbf{1}_N)$)
\begin{eqnarray}
&& \Phi=\mean{\Sigma}=0\,,\\
&& \eta =\log{N\over c}(e^{2m(y-y_0)}+e^{-2m(y-y_0)})\,, 
\label{eta}\\
&& \Sigma^{0} = \frac{1}{2}\partial_y \eta 
= m\tanh \bigl(2m(y-y_0)\bigr)\,, \\
&& H_{L}=\sqrt{\frac{c}{N}}{e^{m(y-y_0)}{\bf 1}_N 
\over \Bigl(e^{2m(y-y_0)}+e^{-2m(y-y_0)}\Bigr)^{1/2}}\,, \\
&& H_{R}=\sqrt{\frac{c}{N}}{e^{-m(y-y_0)}{\bf 1}_N 
\over \Bigl(e^{2m(y-y_0)}+e^{-2m(y-y_0)}\Bigr)^{1/2}}\,.
\end{eqnarray}
Note that in this solution we restore a moduli parameter 
$y_0$ corresponding to the position of the coincident wall. 
A similar construction of domain 
wall solution works for the second sector ($i=2$), 
besides the first sector ($i=1$) given above. 

Let us note that the field-dependent gauge coupling function 
similarly to (\ref{Lag_gChi2}) is automatically obtained 
as a bosonic part of the Lagrangian specified by the 
cubic prepotential in Eq.(\ref{eq:prepotential}), 
Restoring the index $i=1, 2$ for both of the domain wall 
sectors, and by using (\ref{eta}) with 
(\ref{eq:Sigma_by_omega}), 
we finally conclude that the appropriate profile of the 
field-dependent gauge coupling function 
$\Sigma^{0_1}/m_1-\Sigma^{0_2}/m_2$, similarly to 
(\ref{Lag_gChi2}) is achieved. 
When we make (a part of) the global flavor symmetry as a 
local gauge symmetry, we can have several options. 
Since the first flavor group $SU(N_1)$ is generally 
different from the second flavor group $SU(N_2)$, 
we can naturally introduce two 
different gauge fields for the $i=1$ and $2$. 
This option leads to two decoupled sectors in the low-energy 
effective Lagrangian, which can only be coupled by higher 
derivative terms induced by massive modes. 
Another interesting option is to introduce a gauge field 
only for the diagonal subgroup of isomorphic subgroups of two 
different flavor groups, such as 
$SU(\tilde N) \in SU(N_1)$, $SU(\tilde N)\in SU(N_2)$ with 
$\tilde N \le N_1, N_2$. 
This option is interesting in the sense that the massless 
gauge field exchange will communicate between two 
domain wall sectors. 
We hope to come back to these issues in near future.

Let us make a few comments. 
First we have shown that the chiral model analyzed in 
section~\ref{sec:chiral model} can be extended to a 
supersymmetric gauge theory with eight supercharges and 
that the field-dependent gauge coupling function 
which is a clue for localization is naturally explained by 
taking the cubic prepotential. 
Second, there may be more moduli not contained in 
to $(H_{L}^0,H_{R}^0)$, which require further 
studies. Third, here we have presented a solution 
at a special point $\Phi=0$. 
It would be interesting to consider the case or 
$\Phi\neq 0$, but in this case, we need to take a finite 
gauge coupling limit, on which we will investigate in 
future work.

\bigskip
\section{Conclusions and discussion}\label{sc:6}
In this paper we have successfully localized both massless 
non-Abelian gauge fields and massless matter fields 
in non-trivial representation of the gauge group. 
We first considered a (4+1)-dimensional $U(N)$ gauge theory 
with additional $SU(N)_L\times SU(N)_R\times U(1)_{A}$~flavor 
symmetry. 
We introduced the flavor gauge field for the diagonal 
flavor group $SU(N)_{L+R}$, which is unbroken in the 
coincident wall background. 
The flavor gauge fields are localized on the wall by 
introducing the scalar-field-dependent gauge coupling 
function. 
Then we studied the low-energy effective Lagrangian 
and showed that massless localized matter fields 
interact minimally with localized 
$SU(N)_{L+R}$ gauge field as adjoint representations. 
Moreover, full nonlinear interaction between the moduli 
containing up to the second derivatives, was worked out.
The field-dependent gauge coupling function is naturally 
realized in supersymmetric gauge theories 
using the so-called prepotential. 
For this reason, we also explored bosonic part of 
${\cal N}=1$ supersymmetric extension of our model.

Main result of this paper is the effective Lagrangian 
(\ref{eq:full_eff_Lag}). The moduli field $U$ appearing 
in the effective theory, is a chiral $N\times N$ matrix 
field like a pion, since it is a NG boson of spontaneously 
broken chiral symmetry. 
Other moduli in (\ref{eq:full_eff_Lag}), denoted by 
$N\times N$ Hermitian matrix $\hat x$, has the physical 
meaning of positions of $N$ domain walls as its diagonal 
elements. 
We argued that the fluctuations of moduli field $\hat x$, 
can develop VEV corresponding to splitting of walls, 
and the Higgs mechanism will occur as a result.
Namely, the flavor gauge fields get masses by eating 
the non-Abelian cloud. 
Therefore, in this model, Higgs mechanism has a 
geometrical origin like low energy effective theories on D-branes
in superstring theory. 

Amongst the possible future investigations, we would like 
to study non-coincident solution to further clarify this 
geometrical Higgs mechanism. 

We have noticed that our 
effective moduli fields resemble the pion in QCD. 
Similar attempts have been quite successful using 
D-branes \cite{Sakai:2004cn}. 
We believe that our methods can provide more insight 
in various aspects of low-energy hadron physics. 
We plan to explore this direction more fully in 
subsequent studies.

In the discussion of supersymmetric extension of our model 
in section \ref{sec:SUSYE}, we employed a general setup where both sectors 
possessed their own domain wall solution, preserving the 
same half of the supercharges. 
But another alternative approach is also possible. 
We can consider a model, where different halves of 
supercharges are preserved at each sector (BPS and 
anti-BPS walls), and the 
SUSY is completely broken in the system as a whole. 
It has been proposed that the coexistence of BPS and anti-BPS 
walls gives the supersymmetry breaking in a controlled 
manner \cite{Maru:2000sx}. 
In our present case, BPS and anti-BPS sectors interact 
only weakly. 
If we choose flavor gauge field for each sector separately, 
we have only higher derivative interactions induced by 
massive modes. 
If we choose the diagonal subgroup of (subgroups of) 
each sector as flavor gauge group, we have a more 
interesting possibility of the massless gauge field 
as a messenger between two sectors. 
We plan to address this issue elsewhere. 

In order to construct a realistic brane-world scenario 
with the SM fields on the domain wall, we need 
the localization of fields 
in the fundamental representation of the gauge group. 
This is still an open problem and one of the priorities 
of our future investigations. 
In particular, the SM  contains chiral fermions. 
Localization of chiral fermions is a particularly challenging 
problem. 
Anomaly associated with the chiral fermion is also an 
interesting issue to be addressed. 
We would also like to clarify these problems in subsequent 
studies.

Two more issues remain to be addressed. 
 First is the question of sign of gauge kinetic term. 
In our present model, the positivity of the gauge 
coupling function is assured only when positions of 
walls are properly ordered (see 
Eq.(\ref{eq:4d_NA_gauge_coupling})), namely only 
in a region of the moduli space, 
More economical models such as given in Ref.\citen{OhSa} 
may not have such moduli and, therefore, the effective 
gauge coupling may be always positive. 
And lastly, as discussed in section~\ref{sec:SUSYE}, we have not 
succeeded in exhausting all moduli in the supersymmetric 
extension of our model. 
We would also like to investigate these aspects in the future. 

\section*{Acknowledgements} 
This work is supported in part by Japan Society for the 
Promotion of Science (JSPS) and Academy of Sciences of 
the Czech Republic (ASCR) under the Japan - Czech Republic 
Research Cooperative Program, and by Grant-in-Aid for 
Scientific Research from the Ministry of Education, Culture, 
Sports, Science and Technology, Japan No.21540279 (N.S.), 
No.21244036 (N.S.), and No.23740226 (M.E.). 
The work of M.A.~and F.B.~is supported in part by the 
Research Program MSM6840770029 and by the project of 
International Cooperation ATLAS-CERN of the Ministry of 
Education, Youth and Sports of the Czech Republic.

\appendix

\section{Domain walls in the gauged massive $\mathbb{C}P^1$ 
sigma model}\label{app3}

Here we consider the domain wall solutions in the gauged 
massive $\mathbb{C}P^1$ sigma model.
The model is obtained as the strong gauge coupling limit 
of a model similar to that we have studied in 
section~\ref{sec:localization_Abelian}.
Namely, we start with the Lagrangian which has 
$U(1) \times U(1)$ gauge symmetry with two flavors 
\beq
{\cal L} &=& - \frac{1}{4g^2}({\cal F}_{MN})^2 
+ \frac{1}{2g^2}(\p_M\sigma)^2 - \frac{1}{4e^2}
({\cal G}_{MN})^2+ \left|\D_MH\right|^2 - V,\\
V &=& \frac{g^2}{8}\left(|H|^2 - v^2\right)^2 
+ \left|\sigma H - HM\right|^2,
\eeq
where $H=(H_L,H_R)$.
The covariant derivative is given by
\beq
\tilde{\D}_M H = \p_M H + i w_M H + i a_M H q,\quad
q = {\rm diag}(q_L,q_R).
\eeq
The mass matrix is chosen $M = {\rm diag}(m,-m)$ as before.

We next take the strong gauge coupling limit $g\to \infty$ 
of only one of the gauge coupling 
which results in the non-linear sigma model 
coupled to the other gauge field with the finite 
gauge coupling $e$.
In the limit the gauge field $w_M$ and the neutral scalar 
field $\sigma$ become Lagrange multipliers. 
After solving their equations of motion, we have
\beq
w_M = \frac{i}{v^2}\hat\D_M H H^\dagger,\quad
\sigma = \frac{1}{v^2} HMH^\dagger,
\eeq
where we have introduced the covariant derivative
\beq
\hat \D_M H = \p_M H + i a_M H q.
\eeq
Plugging these into the original Lagrangian at 
$g\to \infty$, we get the gauged massive
$\mathbb{C}P^1$ sigma model
\beq
{\cal L}_{g\to\infty} = - \frac{1}{4e^2}
({\cal G}_{MN})^2 + \hat \D_M H P \hat \D^M H^\dagger 
- HMPMH^\dagger,
\label{eq:appC_lag}
\eeq
with the projection operator
\beq
P = 1 - \frac{1}{v^2} H^\dagger H.
\eeq
As before, let us rewrite this Lagrangian with respect 
to the inhomogeneous coordinate
\beq
H = \frac{v}{\sqrt{1+|\phi|^2}}(1,\phi),
\quad \phi \in \mathbb{C}.
\eeq
Then the charge matrix should be chosen as
\beq
q = {\rm diag}(0,1),
\eeq
which leads to a natural expression that the complex 
scalar field $\phi$ has the $U(1)$ charge 1 for the 
gauge field $a_M$:
\beq
\hat \D_M H &=& - \frac{v}{2(1+|\phi|^2)^{3/2}}
\left(\p_M |\phi|^2,\ \phi\p_M|\phi|^2 
- 2(1+|\phi|^2)\hat \D_M\phi\right),\\
\hat \D_M \phi &=& (\p_M + i a_M) \phi.
\eeq
Plugging these into Eq.(\ref{eq:appC_lag}), we finally 
get the Lagrangian
\beq
{\cal L}_{g\to\infty} = - \frac{1}{4e^2}
({\cal G}_{MN})^2 + v^2 
\frac{|\hat \D_M \phi|^2}{(1+|\phi|^2)^2} 
- v^2 \frac{4m^2|\phi|^2}{(1+|\phi|^2)^2}.
\eeq

Let us next consider a domain wall solution in this model. 
We assume all the fields depend on only the 
extra-dimensional coordinate $y$.
Then the four dimensional components of the Maxwell equation
\beq
\p_N {\cal G}^{NM} = i e^2v^2\frac{\hat \D^M\phi \phi^* 
- \phi\hat\D^M\phi^* }{(1+|\phi|^2)^2},
\eeq
can be immediately solved by
\beq
a_\mu = 0,\quad \mu = 0,1,2,3.
\eeq
The fifth component is 
\beq
0 = i e^2v^2\frac{\hat \D^y\phi \phi^* 
- \phi \hat \D^y\phi^*}{(1+|\phi|^2)^2}.
\label{eq:appC_condition}
\eeq
Now the Hamiltonian reduces to the following form
\beq
{\cal H} &=& \frac{v^2}{(1+|\phi|^2)^2}
\left(|\hat \D_y \phi|^2 + 4m^2|\phi|^2\right) \non
&=& \frac{v^2}{(1+|\phi|^2)^2}
\left(|\hat \D_y \phi + 2m\phi |^2 
- 2m \p_y |\phi|^2\right) \non
&\ge& 2mv^2 \frac{d}{dy}\frac{1}{1+|\phi|^2}.
\eeq
Thus the reduced Hamiltonian is minimized when 
the following first order equation is satisfied
\beq
\hat \D_y \phi = - 2m\phi.
\eeq
Since the mass parameter $m$ is real, 
Eq.~(\ref{eq:appC_condition}) is also satisfied.
Let us take the gauge where
\beq
a_y = 0.
\eeq
Then we have the explicit domain wall solution
\beq
\phi = C^2 e^{-2my},\quad C^2 = e^{2i \alpha + 2my_0}.
\label{eq:appC_sol}
\eeq
This is completely the same as the domain wall solution 
given in Eq.~(\ref{eq:sol_u1})
in the ungauged massive $\mathbb{C}P^1$ sigma model.

The final step is to obtain a low energy effective theory 
on the domain wall.
The effective Lagrangian is given by
\begin{align}
 {\cal L}^{\rm eff}_{g\to\infty} 
&= \int dy \left[
- \frac{1}{4e^2}({\cal G}_{\mu\nu})^2 - \frac{1}{2e^2}
({\cal G}_{\mu y})^2 + v^2 \frac{|\hat D_\mu \phi|^2}
{(1+|\phi|^2)^2}
\right] \nonumber \\ 
&= \int dy \left[
- \frac{1}{4e^2}({\cal G}_{\mu\nu})^2 
- \frac{1}{2e^2}({\cal G}_{\mu y})^2 
+ \frac{v^2\left(m^2(\p_\mu y_0)^2
+(\hat\D_\mu \alpha)^2\right)}{\cosh^2 2m(y-y_0)}
\right]\,, \label{eq:appC_eff1}
\end{align}
where we have promoted the moduli parameter $y_0,\alpha$ 
to the fields $y_0(x^\mu)$, $\alpha(x^\mu)$ on the wall,
and we have introduced the covariant derivative
\beq
\hat \D_\mu \alpha = \p_\mu \alpha + \frac{a_\mu}{2},
\eeq
where $\alpha$ is the function of the (3+1)-dimensional 
coordinate $x^\mu$. 
Assuming $a_\mu$ to be y-independent (zero mode), we 
finally obtain 
\begin{equation}
 {\cal L}^{\rm eff}_{g\to\infty} 
= 
- \frac{1}{4e_4^2}({\cal G}_{\mu\nu})^2 
+ \frac{v^2}{m} \left(m^2(\p_\mu y_0)^2
+(\hat\D_\mu \alpha)^2\right) .
\end{equation}
Thus we find that the gauge field $a_\mu(x)$ absorbs 
the scalar field $\alpha(x)$ to become massive via 
the Higgs mechanism. 
Since that the $U(1)$ gauge field $a_\mu$ is massive in 
the effective Lagrangian, we have to 
integrate it out according to the spirit of the low energy 
effective theory.

\bigskip

\section{Effective Lagrangian on the domain wall}\label{app2}

In this appendix we derive our main result (\ref{eq:result}) 
of the effective Lagrangian for the gauged Chiral model 
introduced in \S\ref{sec:chiral model}.

\smallskip
\subsection{Compact form of gauged nonlinear model}

Starting from the Lagrangian using the Einstein summation 
convention for $a=\{L, R\}$ 
\begin{equation}\label{eq:lagr}
\oper{L}_{\rm eff} =  \int_{-\infty}^{\infty} dy\,
\Tr\Bigl[\hat D_{\mu}H_a\hat D^{\mu}H_a^{\dagger}
-v^2W_{\mu}W^{\mu}\Bigr]\,,
\end{equation}
with the constraint 
\begin{equation}\label{eq:constraint}
H_aH_a^{\dagger} = v^2\mathbf{1}_N\,,
\end{equation}
we first eliminate the gauge fields $W_\mu$ to obtain 
a simple expression for gauged nonlinear sigma model. 
Gauge fields $W_{\mu}$ are given by equations of motion as 
\begin{equation}
W_{\mu} = \frac{i}{2v^2}\left[\hat D_{\mu}H_a  
H_a^{\dagger}-H_a \hat D_{\mu} H_a^{\dagger}\right] ,
\end{equation}
and
\begin{equation}
\hat D_{\mu} H = \partial_{\mu}H - i H A_{\mu}\,.
\end{equation} 
The effective Lagrangian \refer{eq:lagr} should also 
contain kinetic term for gauge field $A_{\mu}$, but 
we will not explicitly write it here, for brevity.
Eq. \refer{eq:lagr} can be further simplified by using 
the following identities
\begin{gather}
H_a \hat D_{\mu} H_b^{\dagger} = \partial_{\mu}
(H_aH_b^{\dagger})-\hat D_{\mu}H_a H_b^{\dagger}\,, \\
H_a^{\dagger}\hat D_{\mu} H_b = -\hat D_{\mu} H_a^{\dagger}H_b 
+ \oper{D}_{\mu}H_{ab}\,, 
\end{gather}
where
\begin{equation}
\oper{D}_{\mu} H_{ab} = \partial_{\mu}H_{ab} 
+i\comm{A_{\mu}}{H_{ab}}\,, \hspace{1cm} 
H_{ab} \equiv H_a^{\dagger}H_b\,.
\end{equation}
After some algebra we find:
\begin{equation*}
W_{\mu}W^{\mu} = \tfrac{1}{v^2}\left(\hat D_{\mu} H_a 
\hat D^{\mu}H_a^{\dagger}\right) - 
\tfrac{1}{2v^4}\left(\oper{D}_{\mu}H_{ab}\oper{D}^{\mu}
H_{ba}\right)\,.
\end{equation*} 
Plugging above expression back into the \refer{eq:lagr} 
we arrive at:
\begin{equation}\label{eq:lagr2}
\oper{L}_{\rm eff} 
= \frac{1}{2v^2}\int_{-\infty}^{\infty} dy\, 
\Tr\Bigl[\oper{D}_{\mu}H_{ab}
\oper{D}^{\mu}H_{ba}\Bigr]\,.
\end{equation}

\smallskip
\subsection{Effective Lagrangian}

Now we are ready to compute effective Lagrangian. 
Using a solution (with $\hat y = my\mathbf{1}_N-\hat x$):
\begin{equation}
H =  (H_L, H_R) = \biggl(\frac{v}{\sqrt{2}}
\frac{e^{\hat y/2}}{\sqrt{\cosh(\hat y)}},
\frac{v}{\sqrt{2}}\frac{e^{-\hat y/2}
U^{\dagger}}{\sqrt{\cosh(\hat y)}}\biggr)\,.
\end{equation}
Our new fields $H_{ab}$ are given as:
\begin{align}
H_{LL} & = \frac{v^2}{2}\frac{e^{\hat y}}{\cosh(\hat y)}\,, \\
H_{LR} & = \frac{v^2}{2}\frac{1}{\cosh(\hat y)}U^{\dagger}
=H_{RL}^\dagger\,, 
\\
H_{RR} & = \frac{v^2}{2}U\frac{e^{-\hat y}}{\cosh(\hat y)}
U^{\dagger}\,.
\end{align}

It can be checked, that \refer{eq:lagr2} is given as:
\begin{gather}
\oper{L}_{\rm eff} = \frac{v^2}{4}\int_{-\infty}^{\infty} 
d y\,\Tr\biggl\{\oper{D}_{\mu}
\frac{e^{\hat y}}{\cosh(\hat y)}\oper{D}^{\mu}
\frac{e^{\hat y}}{\cosh(\hat y)}
+\oper{D}_{\mu}\frac{1}{\cosh(\hat y)}\oper{D}^{\mu}
\frac{1}{\cosh(\hat y)} \nonumber \\+
U^{\dagger}\oper{D}_{\mu}U\comm{\frac{e^{-\hat y}}
{\cosh(\hat y)}}
{U^{\dagger}\oper{D}^{\mu}\Bigl(U\frac{e^{-\hat y}}
{\cosh(\hat y)}\Bigr)}
+U^{\dagger}\oper{D}_{\mu}U\comm{\frac{1}
{\cosh(\hat y)}}{\oper{D}^{\mu}\frac{1}
{\cosh(\hat y)}} \nonumber \\+
 \oper{D}_{\mu}U^{\dagger}\oper{D}^{\mu}U
\frac{1}{\cosh^2(\hat y)}\biggr\}\,. \label{master}
\end{gather}
In the following we would like to carry out the integration
over the extra-dimensional coordinate $y$. 
This can be done in two steps. 
First, we must factorize all quantities depending 
on $y$ (or on $\hat y$) to one term inside the trace, 
effectively reducing our problem to fit the following form:
\begin{equation}
\int_{-\infty}^{\infty} dy\, 
\Tr\Bigl[f(my\mathbf{1}_N-\hat x)M\Bigr]\,,
\end{equation} 
where $M$ is some matrix, independent of $y$ and $f$ 
some function. In the second step we diagonalize $\hat x$:
\begin{equation*}
\hat x = P^{-1}\mathrm{diag}(\lambda_1,\ldots ,\lambda_N)P\,,
\end{equation*}
and use the fact that $f(P^{-1}\hat y P)=P^{-1}f(\hat y)P$. 
This transformation leads to
\begin{equation*}
\int_{-\infty}^{\infty} dy\,
\Tr\Bigl[f\bigl(my\mathbf{1}_N
-\mathrm{diag}(\lambda_i)\bigr)PMP^{-1}\Bigr]
= \int_{-\infty}^{\infty} dy\,
\sum_{i=1}^{\lambda}f(my-\lambda_i)(PMP^{-1})_{ii}\,.
\end{equation*}
For every term in the sum we can perform substitution 
$\tilde y = my-\lambda_i$. The key observation is 
that in each term the integration 
will be the same and independent on a particular value 
of $\lambda_i$. Thus we arrive at an identity
\begin{equation}\label{eq:ident}
\int_{-\infty}^{\infty} dy\, \Tr\Bigl[f(\hat y)M\Bigr] 
= \frac{1}{m}\Tr (M)\int_{-\infty}^{\infty} d\tilde y\, 
f(\tilde y)\,.
\end{equation}
It appears as if we just made a substitution 
$\hat y = \tilde y\mathbf{1}_N$. 
This is possible, of course, only thanks to the 
diagonalization trick and properties of the trace. 
In the subsequent subsections, however, we will refer 
to this procedure as 
if it is just a `substitution', for brevity.
Let us decompose the effective Lagrangian \refer{master} 
into three pieces 
\begin{equation}
\oper{L}_{{\rm eff}} = \oper{T}_{\hat x} + \oper{T}_{U}
+\oper{T}_{mixed}
\end{equation}
and see the outlined procedure for each term.

\smallskip
\subsubsection{Kinetic term for $U$}
First, let us concentrate only on terms containing double 
derivatives of $U$, which we denote $\oper{T}_{U}$:
\begin{equation*}
\oper{T}_{U} = \frac{v^2}{4}\int_{-\infty}^{\infty} dy\, 
\Tr\biggl\{\oper{D}_{\mu}U^{\dagger}\oper{D}^{\mu}U
\frac{1}{\cosh^2(\hat y)}
+\oper{D}_{\mu}U^{\dagger}U\comm{\frac{e^{\hat y}}
{\cosh(\hat y)}}{U^{\dagger}\oper{D}^{\mu}U}
\frac{e^{-\hat y}}{\cosh(\hat y)}\biggr\}\,,
\end{equation*}
where we have used the fact that inside commutator 
it is possible to freely interchange
\begin{equation*}
\frac{e^{-\hat y}}{\cosh(\hat y)}\to 
-\frac{e^{\hat y}}{\cosh(\hat y)}\,,
\end{equation*}
since the difference is just a constant matrix. 
In this way we made $\oper{T}_{U}$ manifestly invariant
under exchange $\hat y \to -\hat y$.

Since in the first factor of $\oper{T}_{U}$ all $\hat y$-dependent quantities are on the right side, we can, according to our 
previous discussion, make use of the identity \refer{eq:ident} and carry out the integration:
\begin{gather*}
\frac{v^2}{4}\int_{-\infty}^{\infty} dy\, \Tr\Bigl[\oper{D}_{\mu}U^{\dagger}\oper{D}^{\mu}U\frac{1}{\cosh^2(\hat y)}\Bigr]   
= \frac{v^2}{2m}\Tr\Bigl[\oper{D}_{\mu}U^{\dagger}\oper{D}^{\mu}U\Bigr]\,.
\end{gather*}
For the second term, however, we first use the identity:
\begin{equation}\label{id1}
\comm{f(A)}{B} = \sum_{k=1}^{\infty}\frac{1}{k!}\oper{L}_{A}^k(B)f^{(k)}(A)\,,
\end{equation}
where $\oper{L}_A(B) = \comm{A}{B}$ is a Lie derivative with respect to $A$. Thus
\begin{equation*}
\comm{\frac{e^{\hat y}}{\cosh(\hat y)}}{U^{\dagger}\oper{D}^{\mu}U}=
\sum_{k=1}^{\infty}\frac{(-1)^k}{k!}\oper{L}_{\hat x}^k(U^{\dagger}\oper{D}^{\mu}U)\biggl(\frac{e^{\hat y}}{\cosh(\hat y)}\biggr)^{(k)}\,.
\end{equation*}
Now all $\hat y$-dependent factors are standing on the right and we can formally exchange $\hat y \to \tilde y$. 
The summation can be carried out to get:
\begin{equation}
\sum_{k=1}^{\infty}\frac{(-1)^k}{k!}\oper{L}_{\hat x}^k(U^{\dagger}\oper{D}^{\mu}U)\biggl(\frac{e^{\tilde y}}{\cosh(\tilde y)}\biggr)^{(k)} = 
\frac{e^{\tilde y-\oper{L}_{\hat x}}}{\cosh(\tilde y-\oper{L}_{\hat x})}(U^{\dagger}\oper{D}^{\mu}U)-
\frac{e^{\tilde y}}{\cosh(\tilde y)} U^{\dagger}\oper{D}^{\mu}U\,.
\end{equation}
The formula for $\oper{T}_{U}$ now reads:
\begin{equation}
\oper{T}_{U} = \frac{c}{4m}\int_{-\infty}^{\infty} d\tilde y\, \Tr\biggl[\frac{e^{-\oper{L}_{\hat x}}}{\cosh(\tilde y-\oper{L}_{\hat x})\cosh(\tilde y)}
(U^{\dagger}\oper{D}^{\mu}U)\oper{D}_{\mu}U^{\dagger}U\biggr]\,.
\end{equation}
Since we started with $\oper{T}_{U}$ invariant under the transformation
$\hat y \to -\hat y$, we should take only even part of the above formula (under exchange $\oper{L}_{\hat x} \to -\oper{L}_{\hat x}$) 
as the final result:
\begin{equation}
\oper{T}_{U} = \frac{c}{4m}\int_{-\infty}^{\infty} d\tilde y\, \Tr\biggl[\frac{\cosh(\oper{L}_{\hat x})}{\cosh(\tilde y-\oper{L}_{\hat x})\cosh(\tilde y)}
(U^{\dagger}\oper{D}^{\mu}U)\oper{D}_{\mu}U^{\dagger}U\biggr]\,.
\end{equation}
Now we can carry out the integration using primitive function 
\begin{gather*}
\int\frac{d y }{\cosh(y-\alpha)\cosh(y)} = \frac{1}{\sinh(\alpha)}
\ln\frac{1}{1-\tanh(\alpha)\tanh(y)}\,.
\end{gather*}
Therefore we obtain the result to all orders in $\hat x$ as:
\begin{equation}
\oper{T}_{U} = \frac{v^2}{4m}\Tr\biggl[\oper{D}_{\mu}U^{\dagger}U\frac{1}{\tanh(\oper{L}_{\hat x})}\ln\biggl(\frac{1+\tanh(\oper{L}_{\hat x})}{1-\tanh(\oper{L}_{\hat x})}\biggr)
(U^{\dagger}\oper{D}^{\mu}U)\biggr]\,.
\end{equation}
Performing the Taylor-expansion of the function 
\begin{equation}
 \frac{1}{\tanh(x)}\ln\biggl(\frac{1+\tanh(x)}{1-\tanh(x)}\biggr) = 2+\frac{2 x^2}{3}-\frac{2 x^4}{45}+\frac{4 x^6}{945}-\frac{2 x^8}{4725}+\frac{4 x^{10}}{93555}+O(x^{12}) \; , 
\end{equation}
we can easily read off coefficients of terms beyond the leading one. For example, the first three terms reads:
\begin{multline}
\oper{T}_{U}  = \frac{v^2}{2m}\Tr\Bigl(\oper{D}_{\mu}U^{\dagger}\oper{D}^{\mu}U\Bigr)-\frac{v^2}{6m}
\Tr\Bigl(\comm{\hat x}{U^{\dagger}\oper{D}_{\mu}U}\comm{\hat x}{\oper{D}^{\mu}U^{\dagger}U}\Bigr)\\
-\frac{v^2}{90m}\Tr\Bigl(\comm{\hat x}{\comm{\hat x}{U^{\dagger}\oper{D}_{\mu}U}}\comm{\hat x}{\comm{\hat x}{\oper{D}^{\mu}U^{\dagger}U}}\Bigr)
+\ldots 
\end{multline}

\smallskip
\subsubsection{Mixed term}
Mixed term between $\hat x$ and $U$ is given by
\begin{equation*}
\oper{T}_{mixed} = \frac{v^2}{4}\int_{-\infty}^{\infty} dy\, \Tr\biggl\{U^{\dagger}\oper{D}_{\mu}U\biggl( \comm{\frac{1}{\cosh(\hat y)}}{\oper{D}^{\mu}\frac{1}{\cosh(\hat y)}}
-\comm{\frac{e^{\hat y}}{\cosh(\hat y)}}{\oper{D}^{\mu}\frac{e^{-\hat y}}{\cosh(\hat y)}}\biggr)\biggr\}\,.
\end{equation*}
With use of the identity \refer{id1} and
\begin{equation}
\label{id2} \oper{D}_{\mu}f(\hat x) = \sum_{k=0}^{\infty}\oper{L}_{\hat x}^k(\oper{D}_{\mu}\hat x)\frac{f^{(k+1)}(\hat x)}{(k+1)!}\,,
\end{equation}
one can prove the following:
\begin{equation*}
\label{id3} \comm{f(\hat x)}{\oper{D}_{\mu}g(\hat x)} = \sum_{n=2}^{\infty}\frac{1}{n!}\oper{L}_{\hat x}^{n-1}(\oper{D}_{\mu}\hat x)
\Bigl[\Bigl(f(\hat x)g(\hat x)\Bigr)^{(n)}-f^{(n)}(\hat x)g(\hat x)-f(\hat x)g^{(n)}(\hat x)\Bigr]\,.
\end{equation*}
We can use this result to factorize all $\hat y$-dependent quantities to the right and make the substitution $\hat y = \tilde y\mathbf{1}_N$:
\begin{multline*}
\oper{T}_{mixed} = \frac{v^2}{4m}\int_{-\infty}^{\infty} d\tilde y\, \sum_{n=2}^{\infty}\frac{(-1)^n}{n!}\Tr\Bigl[U^{\dagger}\oper{D}_{\mu}U\oper{L}_{\hat x}^{n-1}
(\oper{D}^{\mu}\hat x)\Bigr]\\ \times\biggl[\biggl(\frac{e^{\tilde y}}{\cosh(\tilde y)}\biggr)^{(n)}\frac{e^{-\tilde y}}{\cosh(\tilde y)}
+\biggl(\frac{e^{-\tilde y}}{\cosh(\tilde y)}\biggr)^{(n)}\frac{e^{\tilde y}}{\cosh(\tilde y)}
-2\biggl(\frac{1}{\cosh(\tilde y)}\biggr)^{(n)}\frac{1}{\cosh(\tilde y)}\biggr]\,.
\end{multline*}
Now we are free to perform summation and integration to obtain:
\begin{equation}
\oper{T}_{mixed} = \frac{v^2}{2m}\Tr\biggl[U^{\dagger}\oper{D}_{\mu}U\,\frac{\cosh(\oper{L}_{\hat x})-1}{\oper{L}_{\hat x}\sinh(\oper{L}_{\hat x})}
\ln\biggl(\frac{1+\tanh(\oper{L}_{\hat x})}{1-\tanh(\oper{L}_{\hat x})}\biggr)(\oper{D}^{\mu}\hat x)\biggr]\,.
\end{equation}
Performing the Taylor-expansion of the function 
\begin{equation}
\frac{\cosh(x)-1}{x\sinh(x)}\ln\biggl(\frac{1+\tanh(x)}{1-\tanh(x)}\biggr) =
x-\frac{x^3}{12}+\frac{x^5}{120}-\frac{17 x^7}{20160}+\frac{31 x^{9}}{362880}+O(x^{11}) ,
\end{equation}
we can easily read off coefficients of terms beyond the 
leading order in the series expansion: 
\begin{equation}
\oper{T}_{mixed} = \frac{v^2}{2m}\Tr\Bigl[U^{\dagger}\oper{D}_{\mu}U\comm{\hat x}{\oper{D}^{\mu}\hat x}\Bigr]
-\frac{v^2}{24m}\Tr\Bigl[U^{\dagger}\oper{D}_{\mu}U\comm{\hat x}{\comm{\hat x}{\comm{\hat x}{\oper{D}^{\mu}\hat x}}}\Bigr]+\ldots 
\end{equation}

\smallskip
\subsubsection{Kinetic term for $\hat x$}

Kinetic term for $\hat x$ is given by
\begin{equation}
\oper{T}_{\hat x} = \frac{v^2}{4}\int_{-\infty}^{\infty} dy\, \Tr\biggl\{
\oper{D}_{\mu}\frac{1}{\cosh(\hat y)}\oper{D}^{\mu}\frac{1}{\cosh(\hat y)}
-\oper{D}_{\mu}\frac{e^{\hat y}}{\cosh(\hat y)}\oper{D}^{\mu}\frac{e^{-\hat y}}{\cosh(\hat y)}
\biggr\}\,.
\end{equation}
We are going to need the identity
\begin{multline}
\label{id4} \Tr\Bigl[\oper{D}_{\mu}f(\hat x)\oper{D}^{\mu}g(\hat x)\Bigr] = \sum_{n=2}^{\infty}\frac{(-1)^n}{n!}\Tr\biggl\{
\oper{L}_{\hat x}^{n-2}(\oper{D}_{\mu}\hat x)\oper{D}^{\mu}\hat x\\\times
\Bigl[\Bigl(f(\hat x)g(\hat x)\Bigr)^{(n)}-f^{(n)}(\hat x)g(\hat x)-f(\hat x)g^{(n)}(\hat x)\Bigr]\biggr\}\,.
\end{multline}
With the aid of this we arrive at
\begin{multline*}
\oper{T}_{\hat x} = \frac{v^2}{4m}\int_{-\infty}^{\infty} d\tilde y\,\sum_{n=2}^{\infty}\frac{1}{n!}
\Tr\Bigl[\oper{L}_{\hat x}^{n-2}(\oper{D}_{\mu}\hat x)\oper{D}^{\mu}\hat x\Bigr] \\
\times\biggl[\biggl(\frac{e^{\tilde y}}{\cosh(\tilde y)}\biggr)^{(n)}\frac{e^{-\tilde y}}{\cosh(\tilde y)}
+\biggl(\frac{e^{-\tilde y}}{\cosh(\tilde y)}\biggr)^{(n)}\frac{e^{\tilde y}}{\cosh(\tilde y)}
-2\biggl(\frac{1}{\cosh(\tilde y)}\biggr)^{(n)}\frac{1}{\cosh(\tilde y)}\biggr]\,,
\end{multline*}
where we again employed diagonalization trick and identity \refer{eq:ident}. Let us carry out the summation and the integration to obtain:
\begin{equation}
\oper{T}_{\hat x} = \frac{v^2}{2m}\Tr\biggl[\oper{D}_{\mu}\hat x\,\frac{\cosh(\oper{L}_{\hat x})-1}{\oper{L}_{\hat x}^2\sinh(\oper{L}_{\hat x})}
\ln\biggl(\frac{1+\tanh(\oper{L}_{\hat x})}{1-\tanh(\oper{L}_{\hat x})}\biggr)(\oper{D}^{\mu}\hat x)\biggr]\,,
\end{equation}
leading to the power series:
\begin{equation}
\oper{T}_{\hat x} = \frac{v^2}{2m}\Tr\Bigl[\oper{D}_{\mu}\hat x\oper{D}^{\mu}\hat x\Bigl]+
\frac{v^2}{24m}\Tr\Bigl[\comm{\hat x}{\oper{D}_{\mu}\hat x}\comm{\hat x}{\oper{D}^{\mu}\hat x}\Bigl]+
\ldots 
\end{equation}

\smallskip

Putting all pieces together as 
$\oper{L}_{{\rm eff}} = \oper{T}_{\hat x} + \oper{T}_{U}
+\oper{T}_{mixed}$, we obtain our final result 
(\ref{eq:result}). 

\bigskip
\section{Determinant of $G$}
\label{app:det}
In order to calculate determinant of matrix $G$ 
\refer{eq:g} we will use the following recurrence formula, 
which relates determinant of 
a symmetric matrix $\oper{M}$ of rank $N+1$ to determinant 
of its $N\times N$ submatrix $M$:
\begin{align}
\label{eq:bigm}\oper{M} & \equiv 
\begin{pmatrix}
 \alpha & u^{T} \\
u & M
\end{pmatrix}\,,  \\ 
\label{eq:detid} \det \oper{M} & = 
(\alpha -u^{T}M^{-1}u)\det M\,.
\end{align}
After double application of formula \refer{eq:detid} we get
\begin{gather}
\det G =  \nonumber \\
\Biggl[\frac{1}{\hat g_1^2}-\bigl(0,\tfrac{\lambda_1}{m_1}
\Phi^{B_1},\tfrac{\lambda_2}{m_1}\Phi^{B_2}\bigr)
\begin{pmatrix}
 & & \\
 & G_1 & \\
 & & 
\end{pmatrix}^{-1}\begin{pmatrix}
0 \\
\tfrac{\lambda_1}{m_1}\Phi^{A_1} \\
\tfrac{\lambda_2}{m_1}\Phi^{A_2}
\end{pmatrix}\Biggr] \nonumber \\
\times \biggl[\frac{1}{\hat g_2^2} 
-\bigl(-\tfrac{\lambda_1}{m_2}
\Phi^{B_1},-\tfrac{\lambda_2}{m_2}\Phi^{B_2}\bigr)
\begin{pmatrix}
 & & \\
 & G_2 & \\
 & &
\end{pmatrix}^{-1}\begin{pmatrix}
-\tfrac{\lambda_1}{m_2}\Phi^{A_1} \\
-\tfrac{\lambda_2}{m_2}\Phi^{A_2}
\end{pmatrix}\biggr] \nonumber \\
\times \Bigl(\frac{1}{e_1^2}\Bigr)^{N_1}\times 
\Bigl(\frac{1}{e_2^2}\Bigr)^{N_2}\,,
\end{gather}
where
\begin{align}
G_1 &= 
\begin{pmatrix}
\frac{1}{\hat g_2^2} & -\frac{\lambda_1}{m_2}\Phi^{A_1} 
& -\frac{\lambda_2}{m_2}\Phi^{A_2} \\
-\frac{\lambda_1}{m_2}\Phi^{B_1} 
& \frac{1}{e_1^2}\delta^{A_1B_1} & 0 \\
-\frac{\lambda_2}{m_2}\Phi^{B_2} & 0 
& \frac{1}{e_2^2}\delta^{A_2B_2}
\end{pmatrix}\,, \\
G_2 &=
\begin{pmatrix}
\frac{1}{e_1^2}\delta^{A_1B_1} & 0 \\
0 & \frac{1}{e_2^2}\delta^{A_2B_2}
\end{pmatrix}\,.
\end{align}
Inverse of $G_1$ is given as 
\begin{gather}
G_{1}^{-1} = \frac{1}{1-\hat g_2^2m_1^2\tilde\Phi^2}\times 
\nonumber \\
\begin{pmatrix}
\hat g_2^2 & e_1m_1\hat g_2^2\tilde\Phi^{A_1} 
& e_2m_1\hat g_2^2\tilde\Phi^{A_2} \\
e_1m_1\hat g_2^2\tilde\Phi^{B_1} 
& e_1^2\delta^{A_1B_1}-m_1^2e_1^2\hat g_2^2\tilde\Phi^{A_1B_1} 
& m_1^2e_1e_2\hat g_2^2\tilde\Phi^{B_1}\tilde\Phi^{A_2} \\
e_2m_1\hat g_2^2\tilde\Phi^{B_2} 
& m_1^2e_1e_2\hat g_2^2\tilde\Phi^{B_2}\tilde\Phi^{A_1} 
& e_2^2\delta^{A_2B_2}-m_1^2e_2^2\hat g_2^2\tilde\Phi^{A_2B_2}
\end{pmatrix}\,,
\end{gather}
where we have used (\ref{eq:tl1}-\ref{eq:tl4}). 

Straightforward calculation leads us to
\begin{equation}
\det G =
\Bigl[\frac{1}{\hat g_1^2}-\frac{m_2^2\tilde\Phi^2}{1-\hat g_2^2m_1^2
\tilde\Phi^{2}}\Bigr]\Bigl[\frac{1}{\hat g_2^2}-m_1^2
\tilde\Phi^2\Bigr]
\Bigl(\frac{1}{e_1^2}\Bigr)^{N_1}\Bigl(\frac{1}{e_2^2}
\Bigr)^{N_2}\,.
\end{equation}
After multiplying both brackets we obtain the result 
\refer{eq:detg}:
\begin{equation}
\det G = \Bigl[\frac{1}{\hat g_1^2\hat g_2^2}
-\left(\frac{m_1^2}{\hat g_1^2}+\frac{m_2^2}{\hat g_2^2}\right)
\tilde\Phi^{2}\Bigr]
\Bigl(\frac{1}{e_1^2}\Bigr)^{N_1}
\Bigl(\frac{1}{e_2^2}\Bigr)^{N_2}\,.
\end{equation}

Next we would like to find condition, which ensures 
positive definiteness of $G$. 
In other words, we require that all eigenvalues of $G$ are 
non-negative.
We can easily turn \refer{eq:detg} into characteristic 
equation by replacing $\hat g_1^{-2},\hat g_2^{-2},e_1^{-2},e_2^{-2}$ 
with $\hat g_1^{-2}-\lambda,\hat g_2^{-2}-\lambda$,\ 
$e_1^{-2}-\lambda,e_2^{-2}-\lambda$. 
However, since $\tilde\Phi^{2}$ consists of terms 
proportional to either 
$e_1^2$ or $e_2^2$ instead of $e_1^{-2}$ or $e_2^{-2}$, 
we should first multiply the term in the square bracket 
by a factor $e_1^{-2}e_2^{-2}$.
Then, after the replacement and denoting 
$\Phi_i^2=\Phi^{A_i}\Phi^{A_i}, i=1,2$, we obtain 
a characteristic equation of the forth order:
\begin{multline}
\biggl[\Bigl(\frac{1}{\hat g_1^2}-\lambda\Bigr)
\Bigl(\frac{1}{\hat g_2^2}-\lambda\Bigr)
\Bigl(\frac{1}{e_1^2}-\lambda\Bigr)
\Bigl(\frac{1}{e_2^2}-\lambda\Bigr) \\
-\Bigl(\frac{m_1^2}{\hat g_1^2}+\frac{m_2^2}{\hat g_2^2}
-\lambda(m_1^2+m_2^2)\Bigr)
\Bigl(\frac{\tilde\Phi^2}{e_1^2e_2^2}
-\lambda\frac{\lambda_1^2\Phi_1^2
+\lambda_2^2\Phi_2^2}{m_1^2m_2^2}\Bigr)\biggr]\,,
\end{multline}
times the factor 
\begin{equation}
\Bigl(\frac{1}{e_1^2}-\lambda\Bigr)^{N_1-1}
\Bigl(\frac{1}{e_2^2}-\lambda\Bigr)^{N_2-1}\,,
\end{equation}
which clearly leads to positive eigenvalues. 
Expanding the brackets we obtain explicit coefficients of
the characteristic polynomial:
\begin{gather}
\lambda^4 -\Bigl(\frac{1}{\hat g_1^2}+\frac{1}{\hat g_2^2}
+\frac{1}{e_1^2}+\frac{1}{e_2^2}\Bigr)\lambda^3 \nonumber \\
+\Bigl(\frac{1}{\hat g_1^2\hat g_2^2}+\frac{1}{\hat g_1^2e_1^2}
+\frac{1}{\hat g_1^2e_2^2}+\frac{1}{\hat g_2^2e_1^2}
+\frac{1}{\hat g_2^2e_2^2}+\frac{1}{e_1^2e_2^2}
-\frac{m_1^2+m_2^2}{m_1^2m_2^2}(\lambda_1^2\Phi_1^2
+\lambda_2^2\Phi_2^2)\Bigr)\lambda^2 \nonumber \\
-\Bigl(\frac{\hat g_1^2+\hat g_2^2+e_1^2+e_2^2}
{\hat g_1^2\hat g_2^2e_1^2e_2^2}
-\tilde g^2\frac{\lambda_1^2\Phi_1^2
+\lambda_2^2\Phi_2^2}{m_1^2m_2^2\hat g_1^2\hat g_2^2}
-\frac{m_1^2+m_2^2}{e_1^2e_2^2}\tilde\Phi^2\Bigr)\lambda 
+\frac{1-\tilde  g^2\tilde\Phi^2}
{\hat g_1^2\hat g_2^2e_1^2e_2^2}\,.
\label{eq:polynom}
\end{gather}

In order to see the non-negativeness of eigenvalues it 
is not necessary to solve the characteristic equation. 
Generally speaking, characteristic 
equation of a real symmetric matrix can be always put into 
the form 
\begin{equation}
(\lambda -\lambda_1)(\lambda-\lambda_2)\ldots 
(\lambda-\lambda_N) = 0\,,
\end{equation} 
where all roots $\lambda_1,\ldots\,\lambda_N$ are 
real numbers. 
Multiplying all parentheses we see that the coefficients 
$c_k$ of characteristic 
polynomial are given by the sum of all possible 
$k$-tuples of $\lambda$'s with alternating sign:
\begin{eqnarray}
&&(\lambda -\lambda_1)(\lambda-\lambda_2)\ldots 
(\lambda-\lambda_N) \nonumber \\
&&= \lambda^N -(\sum_i \lambda_i)\lambda^{N-1}
+\Bigl(\sum_{i>j}\lambda_i\lambda_j\Bigr)\lambda^{N-2}
-\ldots +(-1)^N \lambda_1\ldots \lambda_N \nonumber \\
&&= \sum_{i=0}^{N}(-1)^{i}c_{i}\lambda^{N-i}\,.
\end{eqnarray}
The positivity of all coefficients $c_k$ turns out 
to be equivalent to the positivity of all eigenvalues 
$\lambda_k$. 
To ensure positivity of the eigenvalues, we now 
demand that all terms in \refer{eq:polynom} inside 
brackets are positive. This gives us three conditions:
\begin{align}
\hat g_1^2\hat g_2^2+\hat g_1^2e_1^2+\hat g_1^2e_2^2+\hat g_2^2e_1^2+\hat g_2^2e_2^2
+e_1^2e_2^2 & \nonumber \\
-\hat g_1^2\hat g_2^2(m_1^2+m_2^2)(e_1^2 \tilde\Phi_2^2
+e_2^2\tilde\Phi_1^2) & \geq 0\,, \\
\hat g_1^2+\hat g_2^2+e_1^2+e_2^2-\tilde g^2 (e_1^2\tilde\Phi_2^2
+e_2^2\tilde\Phi_1^2)-\hat g_1^2\hat g_2^2(m_1^2+m_2^2)
\tilde\Phi^2 & \geq 0\,, \\
1-\tilde  g^2 \tilde\Phi^2 & \geq 0\,.
\end{align}
These can be put into the convenient form:
\begin{align}
1-\tilde g_{21}^2\tilde\Phi_1^2
-\tilde g_{22}^2\tilde\Phi_2^2 &\geq 0\,, \\
1-\tilde g_{11}^2\tilde\Phi_1^2
-\tilde g_{12}^2\tilde\Phi_2^2 &\geq 0\,, \\
1-\tilde g^2\tilde\Phi_1^2
-\tilde g^2\tilde\Phi_2^2 &\geq 0\,,
\end{align}
where 
\begin{align}
\tilde g_{21}^2 & = \frac{\hat g_1^2\hat g_2^2(m_1^2+m_2^2)e_2^2}
{\hat g_1^2\hat g_2^2+\hat g_1^2e_1^2+\hat g_1^2e_2^2+\hat g_2^2e_1^2+\hat g_2^2e_2^2
+e_1^2e_2^2}\,, \\
\tilde g_{22}^2 & = \frac{\hat g_1^2\hat g_2^2(m_1^2+m_2^2)e_1^2}
{\hat g_1^2\hat g_2^2+\hat g_1^2e_1^2+\hat g_1^2e_2^2+\hat g_2^2e_1^2+\hat g_2^2e_2^2
+e_1^2e_2^2}\,, \\
\tilde g_{11}^2 & = \frac{\tilde g^2e_2^2
+\hat g_1^2\hat g_2^2(m_1^2+m_2^2)}{\hat g_1^2+\hat g_2^2+e_1^2+e_2^2}\,, \\ 
\tilde g_{12}^2 & = \frac{\tilde g^2e_1^2
+\hat g_1^2\hat g_2^2(m_1^2+m_2^2)}{\hat g_1^2+\hat g_2^2+e_1^2+e_2^2}\,.
\end{align}
We are going to argue, that the last condition 
$1-\tilde g^2\tilde\Phi^2 \geq 0$ is the strongest one 
and, therefore, only one important.
This can be true if and only if parameter $\tilde g^2$ 
is always  greater then $\tilde g_{11}^2$, 
$\tilde g_{12}^2$, $\tilde g_{21}^2$
and $\tilde g_{22}^2$ for all possible values of 
involved parameters. 
This is indeed so. 
Let us demonstrate this fact by showing, for example
\begin{equation}
\tilde g^2 \geq \tilde g_{11}^2\,.
\end{equation}   
Multiplying both sides by $\hat g_1^2+\hat g_2^2+e_1^2+e_2^2$ 
and expanding our notation, we get
\begin{gather}
(m_1^2\hat g_2^2+m_2^2\hat g_1^2)(\hat g_1^2+\hat g_2^2+e_1^2+e_2^2) 
\geq e_2^2(m_1^2\hat g_2^2+m_2^2\hat g_1^2)+\hat g_1^2\hat g_2^2(m_1^2+m_2^2), 
\end{gather}
leading to
\begin{gather}
m_1^2\hat g_2^2(\hat g_2^2+e_1^2)+m_2^2\hat g_1^2(\hat g_1^2+e_1^2) \geq 0\,.
\end{gather}
Last line is obviously always true. In the same way, 
one can show that $\tilde g^2 \geq \tilde g_{12}^2$,
$\tilde g^2 \geq \tilde g_{21}^2$ and 
$\tilde g^2 \geq \tilde g_{22}^2$.
This proves our claim, that 
condition \refer{detG} is both necessary and 
sufficient to ensure positivity of the matrix $G$.

\end{document}